\documentclass[journal]{IEEEtran}

\IEEEoverridecommandlockouts

\usepackage{relsize}
\usepackage{moreverb}
\usepackage{mathrsfs}
\usepackage{amsmath}
\usepackage{amssymb}
\usepackage{stfloats}
\usepackage{graphicx}
\usepackage{makecell}
\usepackage{xcolor}
\usepackage{cite}
\usepackage[ruled,linesnumbered]{algorithm2e}
\usepackage{multirow}       
\usepackage{booktabs}
\usepackage{multirow}
\usepackage{tabularx,ragged2e,booktabs,caption}

\ifCLASSOPTIONcompsoc
\usepackage[tight,normalsize,sf,SF]{subfigure}
\else
\usepackage[tight,footnotesize]{subfigure}
\usepackage{subfigure}
\usepackage{bm}

\usepackage{amsmath}

\usepackage{setspace}

\begin{document}
\title{Waveform Design for Partial-Time Superimposed ISAC Systems
\thanks{This work was supported in part by the National Key Research and Development Program under Grant 2024YFE0213600; in part by the National Natural Science Foundation of China under Grant 62401473; in part by the National Key Laboratory Fund Project for Space Microwave Communication under Grant HTKJ2024KL504010, in part by the Shenzhen Science and Technology Program under Grant JCYJ20240813150735045; in part by the Key Research and Development Project in Shaanxi Province under Grant 2025CY-YBXM-055 and 2025CG-GJHX-13, in part by Stable Support Program for National Key Laboratory on Aircraft Integrated Flight Control under Grant WDZC2025601B09, in part by  Open Fund Project of the Laboratory for Space Trusted Computing and Electronic Information Technology under Grant OBCandETL-2024-04(02), and in part by the Project NEURONAS which is implemented in the framework of H.F.R.I call "3rd Call for H.F.R.I.’s Research Projects to Support Faculty Members \& Researchers" (H.F.R.I. Project Number: 25726). \emph{(Corresponding author: Rugui Yao and Ye Fan)}}
\thanks{
Xi Nan, Rugui Yao, Ye Fan, Xiaoya Zuo are all with the School of Electronics and Information, Northwestern Polytechnical University, Xi'an, Shaanxi China. Y. Fan is also with the Shenzhen Research Institute of Northwestern Polytechnical University, Room 2501, No. 45, Gaoxin South 9th Road, Nanshan District, Shenzhen, Guangdong 518057, China. (e-mail: nanxi@mail.nwpu.edu.cn; yaorg@nwpu.edu.cn; fanye@nwpu.edu.cn;  zuoxy@nwpu.edu.cn).
}
\thanks{
Ruikang Zhong is with the School of Information and Communication Engineering, Xi'an Jiaotong University.  (e-mail: Ruikang.zhong@xjtu.edu.cn;)
}
\thanks{
Theodoros A. Tsiftsis is with Department of Electrical and Electronic Engineering, University of Nottingham Ningbo China, Ningbo 315100, China, and also with the Department of Informatics and Telecommunications, University of Thessaly, Lamia 35100, Greece. (e-mail: theodoros.tsiftsis@nottingham.edu.cn; tsiftsis@uth.gr).
}
\thanks{Alexandros-Apostolos A. Boulogeorgos is with the Department of Electrical and Computer Engineering, Democritus University of Thrace, 67100 Xanthi, Greece (e-mail: al.boulogeorgos@ieee.org).}
}

\author{\IEEEauthorblockN{Xi Nan,~Rugui Yao,~Ye Fan,~Ruikang Zhong,~Xiaoya Zuo,~Theodoros A. Tsiftsis \\and ~Alexandros-Apostolos A. Boulogeorgos}
%
}
	
\maketitle
	
\begin{abstract}
Nowadays, waveforms of integrated sensing and communication (ISAC) are almost based on conventional  communication and sensing signal, which bounds both the communication and sensing  performance. To deal with this issue, in this paper, a novel waveform design is presented for the partial-time superimposed (PTS) ISAC system. At the base station (BS), a parameter-adjustable linear frequency modulation (LFM) pulse signal and a continuous communication orthogonal frequency division multiplexing (OFDM) signal are employed to broadcast public information and perform sensing tasks, respectively, using a PTS scheme.
Pulse compression gain enhances the system's long-range sensing capability, while OFDM ensures the system's high-speed data transmission capability. Meanwhile, the LFM signal is utilized as superimposed pilot for channel estimation, which has higher time-frequency resource utilization and stronger real-time performance compared to orthogonal pilots. We present an accurate parameter estimation method of multi-path sensing signal for reconstructing and interference cancellation in communication users. Additionally, a cyclic maximum likelihood method is introduced for channel estimation and the Cram\'er-Rao lower bound (CRLB) of channel estimation is derived. Simulations demonstrate the accuracy and robustness of the proposed parameter estimation algorithm as well as the improved channel estimation performance over traditional methods. The proposed waveform design method can achieve reliable data transmission and accurate target sensing.

\end{abstract}

	
\IEEEpeerreviewmaketitle

\section{Introduction}
\IEEEPARstart{T}{he} rapid proliferation of wireless communication devices driven by fifth-generation (5G) technologies and internet of things (IoT) applications have significantly strained available wireless spectrum resource \cite{1}. As a response to the spectrum scarcity, researchers have turned their attention
to spectrum sharing approaches\cite{2}. Integrated sensing and communication (ISAC) enables a waveform to have both communication and sensing functions, which effectively improves spectrum and energy efficiencies as well as reducing the hardware cost. Fortunately, sensing and communication systems become increasingly similar in terms of hardware architectures, channel characteristics, and signal processing, which lays the foundation of the integrated design of communication and sensing systems \cite{3,4,5}. High-precision sensing is considered a key feature of sixth-generation (6G) wireless communication systems as it fulfills the demands of several application scenarios, including extended reality, self-driven vehicles, robotics, etc. As a result,  ISAC design is an important direction of 6G research \cite{6}.

\subsection{Literature review} 

Multiple access technology represents one of the
simplest and most effective approaches to implementing ISAC, enabling communication
and sensing functions through orthogonal resource allocation. Existing studies have
proposed various schemes based on orthogonal resource allocation, including time-division
\cite{7}, frequency-division \cite{8}, code-division \cite{9}, and space-division \cite{10}. However,
these orthogonal multiple access methods isolate communication and sensing functions
across time, frequency, code, or spatial domains, resulting in degraded resource
utilization efficiency.

Another approach is to design integrated waveform
that shares the same resources. And the waveform design primarily follows two paradigms:
communication-centric and sensing-centric. Communication-centric
approaches typically leverage OFDM signals, which are widely adopted in modern wireless
systems due to their high spectral efficiency and compatibility with existing communication
infrastructure. In these schemes, sensing functionality is achieved by processing
the channel information matrix in the time-frequency domain and transforming it
into the delay-Doppler domain using two-dimensional discrete Fourier transform (2D-DFT).
By identifying spectral peaks in the periodogram, the delay and Doppler information
can be estimated, enabling target distance and velocity measurements \cite{11,12,13,14}. However,
OFDM-based ISAC signals often suffer from high peak-to-average power ratio (PAPR),
which can degrade sensing performance, and their sensing range is limited by the
cyclic prefix duration \cite{13,14}. To address these issues, advanced designs such as
null-space precoding, diagonal resource allocation, and robust reference signal
patterns have been proposed to enhance sensing accuracy and reduce ambiguity without
sacrificing communication quality. On the other hand, sensing-centric designs use
typical radar waveforms (such as LFM) as the basis and embed communication data
onto the sensing signal. Techniques like amplitude-shift keying, frequency-shift
keying, phase-shift keying, or modulating communication symbols onto the chirp rate
of LFM have been explored \cite{15,16,17}. These methods can achieve high-resolution sensing
and are robust to hardware nonlinearity, but the achievable communication rate is
generally limited compared to OFDM-based schemes. 

In recent years, non-orthogonal multiple access (NOMA)-enabled
ISAC systems have emerged as a key research direction for 6G and future wireless
networks. Extensive studies have systematically explored the architecture, resource
allocation, interference management, and performance optimization of NOMA-ISAC systems.
NOMA-ISAC enables the efficient coexistence of communication and sensing on shared
spectrum and hardware platforms, significantly improving spectral efficiency and
overall system performance. Various NOMA-ISAC frameworks have been proposed, incorporating
joint beamforming, power allocation, and sensing scheduling optimization algorithms
to simultaneously enhance communication rates and sensing accuracy \cite{18,19,20,21,22,23,24,25,26,27}.
The integration of reconfigurable intelligent surfaces (RIS) and simultaneously
transmitting and reflecting RIS (STAR-RIS) further extends ISAC coverage and sensing
capabilities, enhancing system flexibility and robustness \cite{26,27}. Moreover, NOMA-ISAC
has been combined with multi-tier computing and semi-integrated sensing and communication
(Semi-ISAC) architectures, promoting multi-functionality and efficient resource
utilization \cite{28,29}. Theoretical analyses and simulation results consistently demonstrate that NOMA-ISAC outperforms conventional orthogonal schemes, especially in highly
correlated channels and overloaded scenarios \cite{18,20,25}. Overall, NOMA-ISAC exhibits
significant advantages in some fields.

NOMA improves the spectral efficiency, while, as a communication multi-user access enabler, it comes with some shortcomings that origins due to (i) multi-user interference, (ii) high channel quality requirements, and (iii) erroneous propagation \cite{30}. The aforementioned shortcomings can be translated into the following question: \emph{(i) How to utilize non-orthogonal superposition to enable the complete sharing of time-frequency resources between superior communication and sensing signals, forming a new ISAC signal? (ii) How to separate communication and sensing signals in multipath propagation environments and reduce mutual interference? (iii) And how to achieve mutual assistance between communication and sensing to enhance cooperative gains.}

\subsection{Motivation \& contribution}

Fortunately, we have found a compromised scheme: The sensing signal is partial-time superimposed (PTS) with the communication signal. The communication signal adopts OFDM  to ensure high-speed and robust information transmission, and the sensing signal adopts LFM to guarantee high-precision target sensing. At the same time, the sensing signal is used as the superposed pilot for the channel estimation of the communication user. The mean advantages of the proposed PTS-ISAC waveform are presented as: 1) High time-frequency resource utilization, the sensing signal serves as a superimposed pilot, enabling the reuse of time-frequency resources with the communication signal. Compared to traditional orthogonal pilot designs, this approach achieves higher resource utilization. 2) More flexible waveform design strategy, by dynamically adjusting parameters such as communication-sensing power allocation and the duty cycle of the sensing signal, the ISAC waveform can achieve stronger adaptability in complex communication scenarios and various sensing tasks. 3) Enhanced communication-sensing coordination, PTS-ISAC realizes sensing-assisted communication channel estimation, improving the coordination between the two functions, one of the original visions of ISAC systems.

Additionally, the main contributions of this paper are as follows

\begin{enumerate}
	
	\item We present a PTS ISAC waveform scheme. LFM pulse signal and continuous OFDM signal are superimposed, sharing all the frequency and partial time resource. Pulse compression increases the received SINR of sensing system, while OFDM ensures the system's high-speed data transmission capability. Meanwhile, the LFM signal is also utilized as superimposed pilot for channel estimation, which has higher time-frequency resource utilization. It is worth noting that instead of interfering each other, sensing-assisted communication is achieved in this paper.
	
	\item  An effective multi-path LFM signal parameter detection algorithm is proposed to reconstruct the sensing signal accurately and cancel its interference at the communication receiver. In more detail, a LFM parameter estimation method based on  {blind source separation} (\textsc{BSS}), which involves rough estimation, fine estimation and revision (RFR) is reported. Comparing to the existing algorithm, the accuracy of parameter estimation is obviously improved, especially in the low SNR environment.
	
	\item A sensing-assisted channel estimation method is proposed. The sensing signal is regarded as a superimposed pilot to achieve channel estimation by a proposed cyclic maximum likelihood (CML) method without any additional time or frequency resources. Further, the Cram\'er-Rao lower bound (CRLB) of the channel estimation is derived. 
	
\end{enumerate}

\begin{figure*}[t] 
	\centering
	\begin{minipage}[t]{0.49\textwidth}
		\centering
		\includegraphics[width=\linewidth]{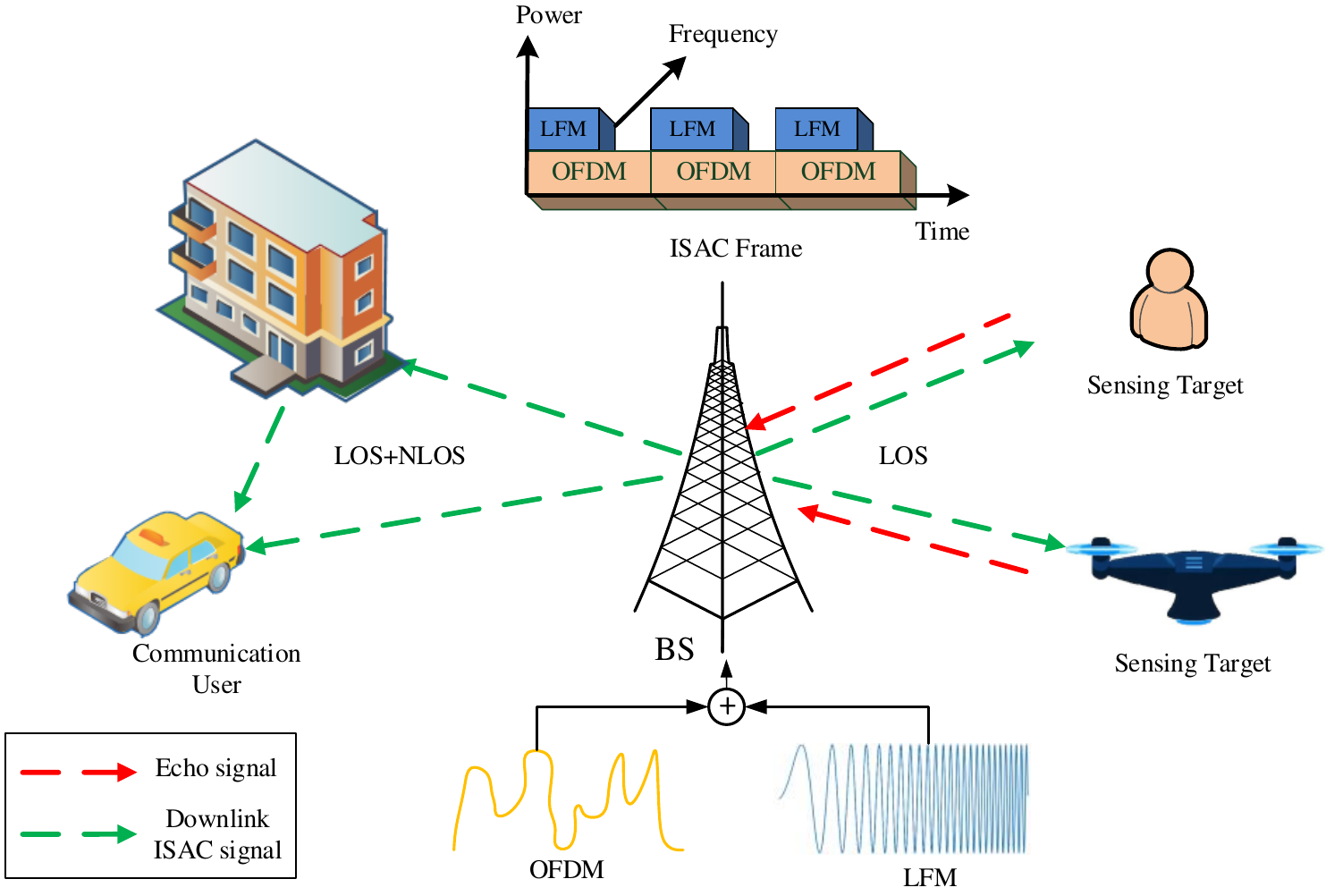} 
		\caption{System model.}
		\label{fig:a}
	\end{minipage}
	\hfill
	\begin{minipage}[t]{0.49\textwidth}
		\centering
		\includegraphics[width=\linewidth]{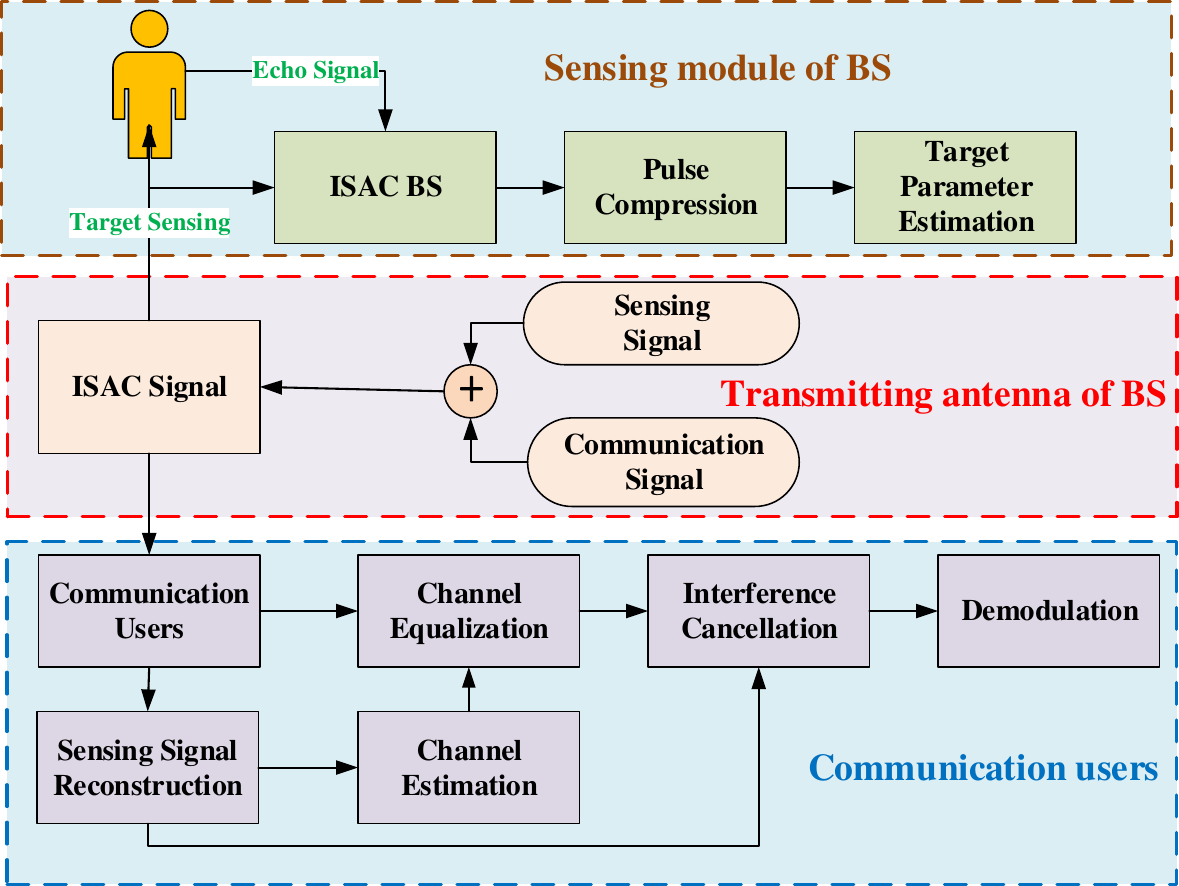} 
		\caption{Signal processing flow chart}
		\label{fig:b}
	\end{minipage}
	\hfill
\end{figure*}

\subsection{Organization \& notations }
The rest of this paper is organized as follows. Section II introduces the system model. In Section III, we propose the parameter estimation method of the sensing signal. And in Section IV, the sensing-assisted communication channel estimation method is proposed. Simulation results are presented in Section V. Finally, Section VI provides concluding remarks.

\emph{Notation:} Throughout this paper, ${\bf{A}}$,  ${\bf{a}}$, $a$ denote matrix, vector, and scalar, respectively. ${{(\cdot )}^{*}}$,${\left(  \cdot  \right)^{\rm{T}}},{\left(  \cdot  \right)^{\rm{H}}},{\left(  \cdot  \right)^{ - 1}},\left|  \cdot  \right|,\mathbb{E}( \cdot )$ ${\rm{trace}}( \cdot )$, $ \otimes $ denote conjugate, transpose, Hermitian, inverse, modulus of a complex number, expectation, trace of a matrix and convolution operation  respectively. Finally, ${{\Bbb C}^{N \times M}}$ denotes the complex space of dimension $N \times M$.

\section{System Model}
We consider a post-disaster ISAC emergency downlink communication scenario \footnote{ Based on the channel reciprocity, the downlink system model established in this paper is applicable to the uplink scenario. The proposed ISAC waveform and the derivations for signal processing and channel estimation remain valid for a single-antenna user accessing a multi-antenna BS.}, the single-antenna ISAC BS broadcasts public information to the multi-antenna user (such as emergency communication vehicle) and search for survivors, which is illustrated in Fig.1. Due to the lack of users' prior information, the ISAC BS omnidirectionally broadcasts the common PTS ISAC signal without precoding to the communication users with $N_a$ antennas through multi-path channels, while sensing multiple targets simultaneously.  The PTS ISAC signal is generated by  weighted superimposing of OFDM signal with public information and LFM signal, which is formulated as
\begin{equation}
	x(t){\rm{ }} = \sqrt {1 - w}   s(t) + \sqrt w   c(t),
\end{equation}
where $s(t)$ is the LFM signal for sensing, and $c(t)$ is the OFDM signal of communication component, $ w $ is the power weighting coefficient. Based the sensing requirements, BS dynamically adjusts the parameters of the LFM signal. In order to improve time-frequency resource utilization, BS does not send additional pilots. Users employ multiple antennas for LFM signal parameters estimation, reconstruction and interference elimination. Moreover, LFM signal is used as a superimposed pilot for channel estimation to demodulate public information for users. We assume that the length of LFM signal is longer than the channel impulse response. The specific signal processing flow is shown in Fig.2.  

\subsection{Communication model}   
OFDM is widely used in various communication systems because of its high frequency spectrum utilization efficiency and excellent anti-multipath performance. Let ${{\mathbf{C}}_{m}}=[{{C}_{m}}[0],{{C}_{m}}[1],\ldots ,{{C}_{m}}[{{N}_{c}}-1]]$, $m=0,1,\ldots ,M-1$ be the discrete frequency domain symbol of the $m$-th OFDM symbol. $M$ stands for the number of OFDM symbol, $N_c$ represents the number of subcarrier. The discrete time domain symbol of the $m$-th OFDM symbol can be obtained by inverse discrete Fourier transform (IDFT) of ${\mathbf{C}}_{m}$ as

\begin{equation}
	\begin{aligned}
		{{c}_{m}}[n] &= \frac{1}{\sqrt{{{N}_{c}}}} 
		\sum\limits_{k=0}^{{{N}_{c}}-1} {{C}_{m}}[k] 
		{{\mathrm{e}}^{2\mathrm{j}\pi \frac{kn}{{{N}_{c}}}}} , \\
		& \qquad \rm{with}\ \  n = 0,1,\ldots ,{{N}_{c}}-1.
	\end{aligned}
\end{equation}

The continuous time domain OFDM signal can be expressed as
\begin{equation}
	c(t) = \frac{1}{{\sqrt {{N_c}} }}\sum\limits_{m = 0}^{M - 1} {\sum\limits_{k = 0}^{{N_c} - 1} {{C_m}[k]  {{\rm{e}}^{{\rm 2j\pi} k\Delta f(t - m{T_c})}}} }   {\Pi}\left( {\frac{{t - m{T_c}}}{{{T_c}}}} \right),
\end{equation}
with
\begin{equation}
	\Pi(t) = 
	\begin{cases} 
		1, & \text{if } 0<t \leq 1 \\
		0, & \text{otherwise}
	\end{cases},
\end{equation}
where ${T_c}$ is the duration of OFDM symbol, $\Delta f = 1/{T_{\rm{c}}}$ is the interval of subcarrier. For simplicity of expression, we omit the user index. And the discrete time domain received signal of $i$-th antenna of user corresponding to the $m$-th OFDM symbol can be expressed as
\begin{equation}
	{{y}_{m}^i}[n]=\sum\limits_{l=0}^{L-1}{h_i[l]{{c}_{m}}[n-l]}+z_i[n], \text{ with }n=0,1,\ldots ,{{N}_{c}}-1,
\end{equation}
where $h_i[l],l=0,1,\ldots ,L-1$ is the channel impulse response and $z_i[n]$ is the Gaussian white noise. The discrete received signal in frequency domain can be obtain by discrete Fourier transform (DFT) of $\left\{ {{y}_{m}^i}[n] \right\}_{n=0}^{{{N}_{c}}-1}$ as
\begin{equation}
	{{Y}_{m}^i}[k]=H_i[k]{{C}_{m}}[k]+Z_i[k], \text{ with }k = 0,1, \ldots ,{N_c} - 1,
\end{equation}
where $\left\{ H_i[k] \right\}_{k=0}^{{{N}_{c}}-1}$ is the DFT of $\left\{ h_i[l] \right\}_{l=0}^{L-1}$. The estimation of ${{C}_{m}}[k]$ by zero forcing equalization can be written as

\begin{equation}
	{{\hat{C}}_{m}}[k]=W[k](H_i[k]{{C}_{m}}[k]+Z_i[k]),\text{ with } k=0,1,\ldots ,{{N}_{c}}-1,
\end{equation}
with
\begin{equation}
	W[k] = \frac{1}{{H_i[k]}},k = 0,1, \ldots ,{N_c} - 1,
\end{equation}
\subsection{Sensing model}

LFM signal is widely used in radar detection due to its large time-bandwidth product and high pulse compression ratio. The signal of LFM pulse radar can be expressed as:
\begin{equation}
	\begin{split}
		s(t) = \sum_{m = 0}^{M - 1} & A\exp \left( \text{j}2\pi \left( f_0(t - m{T_c}) + \frac{1}{2}k(t - m{T_c})^2 \right) \right) \\
		& \times \Pi \left( \frac{t - m{T_c}}{T_s} \right), \quad \text{with } T_s < T_c
	\end{split}
\end{equation}
where ${f_0}$ is the initial frequency, $k$ is the chirp rate, $A$ is the amplitude of the LFM signal, ${T_s}$ is the pulse width, and $M$ is the number of LFM pulses. Setting different ${T_s}$  for the sensing signal essentially corresponds to adjusting the duty cycle. The fundamental significance of this adjustment lies in strategically allocating the sensing's attention and resources based on core mission requirements, so as to optimize its performance across three key dimensions: range, velocity, and unambiguous field of view. 

The LFM signal is non-orthogonal weighted superimposed with the OFDM signal; therefore, the real sensing signal can be written as
\begin{equation}
	{x_r}(t) = {\Pi}\left( {\frac{{t - n{T_c}}}{{{T_s}}}} \right)  x(t).
\end{equation}

The parameter estimation and channel estimation are all based sensing signal,  we just analyze ${x_r}(t)$. We assume that there exist $N_s$ sensing targets. Then, the echo signal can be expressed as
\begin{equation}
	{y_{{\rm e},m}}(t) = \sum\limits_{n = 1}^{{N_s}} {{x_{r,m}}(t - {t_n}){e^{{\rm j2\pi} {f_{d,n}}(t - {t_n})}}} ,
\end{equation}
where $v_n$ and $d_n$ are the velocity and distance of the $n$-th target, respectively,  while ${t_n} = 2{d_n}/c$, ${f_{d,n}} = 2{v_n}{f_0}/c$ are the delay and Doppler frequency shift caused by $n$-th sensing target, respectively. 

The echo signal is forwarded to the pulse compressor. Pulse compression (PC) improves the SNR and spatial resolution, and PC of $m$-th echo signal can be formulated as
\begin{equation}
	\begin{aligned}
		{y_{{\rm pc},m}}(t) &= x_r^ * ( - t) \otimes {y_{{\rm echo},m}}(t).
	\end{aligned}
\end{equation}
After sampling, ${y_{{\rm pc},m}}(t)$ are organized in the following matrix:
\begin{equation}
	{{\mathbf{Y}}_{pc}} = \left| {\begin{array}{*{20}{c}}
			{{y_{{\text{pc}},0}}[0]}&{{y_{{\text{pc}},0}}[1]}& \cdots &{{y_{{\text{pc}},0}}[{N_p} - 1]} \\ 
			{{y_{{\text{pc}},1}}[0]}&{{y_{{\text{pc}},1}}[1]}& \cdots &{{y_{{\text{pc}},1}}[{N_p} - 1]} \\ 
			\vdots & \vdots & \ddots & \vdots  \\ 
			{{y_{{\text{pc}},M - 1}}[0]}&{{y_{{\text{pc}},M - 1}}[1]}& \cdots &{{y_{{\text{pc}},M - 1}}[{N_p} - 1]} 
	\end{array}} \right|,
\end{equation}
where ${N_{p}}$ is the number of samples. By applying DFT to each column of ${{\bf{Y}}_{pc}}$ and searching the peaks, we obtain the estimated values of the targets' speed and distance.

In addition, ambiguity function is an important tool for radar waveform design and analysis. It  characterizes the waveform and the corresponding matched filter. By analyzing the ambiguity function of radar transmitted waveform, the resolution, measurement accuracy and ambiguity of radar system can be obtained, when the optimal matching filter is applied. The definition of ambiguity function adopted in is 

\begin{equation}
	\chi (\tau ,{f_d})=\int_{ - \infty }^{ + \infty } {{x_r}(t) \ x_r^*(t - \tau )  } {{\rm{e}}^{{\rm{2j\pi }}{f_d}t}}{\rm dt},\\
\end{equation}
where $\tau $ stands for the time delay, ${f_d}$ denotes the Doppler frequency shift. A random OFDM communication signal is superimposed on the LFM signal, which reduces its auto-correlation property. This affects the performance of the ambiguity function.

\section{parameter estimation of sensing signal}
The multi-path ISAC signal received at the communication receiver is used to estimate the parameter of the sensing signal and reconstruct the sensing signal, in order to carry out channel estimation and interference cancellation. However, the traditional parameter estimation algorithms of LFM signals considered the multi-component problem but not the multi-path problem \cite{31}. There are also few studies on parameter estimation of multi-path LFM signals, such as the cyclic correlation transformation (CCT) mentioned in \cite{32}. However, its accuracy is limited, especially in the low SNR scenario.

\begin{figure}[t]
	\centering
	\includegraphics[width=9cm]{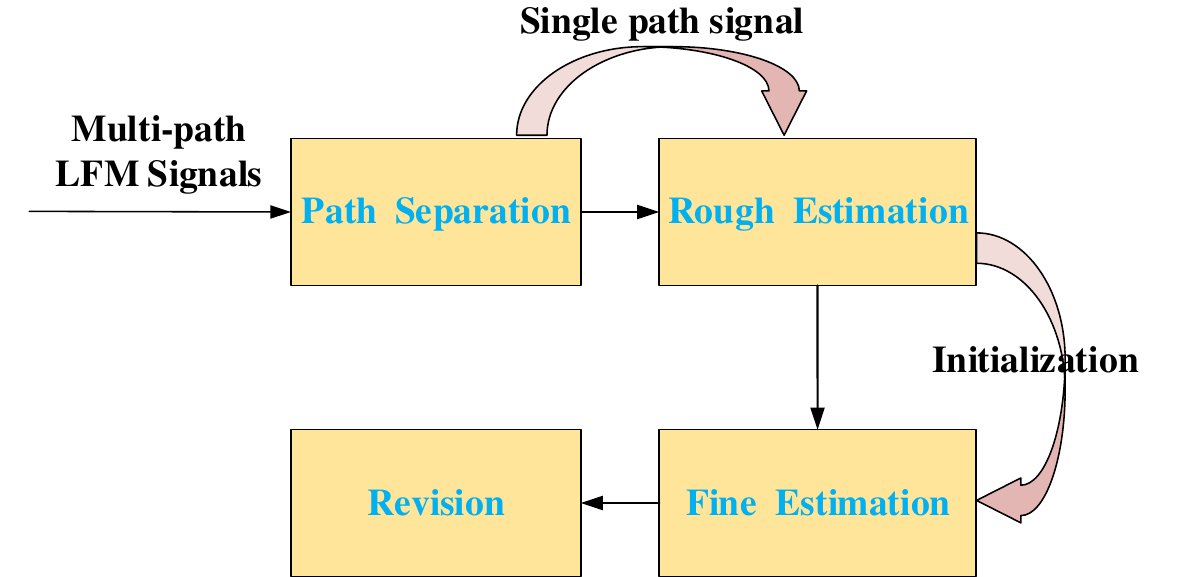}
	\caption{Multi-path LFM signals parameters estimation process}
	\label{fig3}
\end{figure}

We present a multi-path LFM signal parameter estimation algorithm. The BSS method is used to separate the multi-path LFM signals and extracted its single-path signal. Then the short-time Fourier transform (STFT) is employed for rough parameter estimation, and fractional Fourier transform (FRFT) is used for fine parameter estimation. Finally, the simulated annealing (SA) algorithm is used to revise estimated parameters. The parameters estimation process is shown in Fig.~\ref{fig3}.

\subsection{Separation of multi-path LFM signals}
The received multi-path signal is formed with $N_{\rm r}$ ISAC signals with different delays. According to the Central Limit Theorem,  when the number of subcarriers  is sufficiently large, the OFDM signal, generated as the sum of the independent and identically distributed (i.i.d.) subcarriers, converges to a complex Gaussian random variable \cite{33}, which can be regarded as noise. Thereby,  the received signal is formed with $N_{\rm r}$ LFM signals with different delays and noise. We assume that the number of user's antennas $N_{\rm a} \geqslant N_{\rm r}$  have low correlation to each others. As a result, the received signals of $N_{\rm r}$ antennas constitute the observation matrix, which can be expressed as

\begin{equation}
	\begin{split}
		\begin{array}{c}
			{\bf{Y}} = \left[ {{{\bf{y}}_1},{{\bf{y}}_2}, \ldots ,{{\bf{y}}_{N_{\rm r}}}} \right]^{\rm T}
			= {\bf{MS}} + {\bf{Z}},
		\end{array}
	\end{split}
\end{equation}
where
\begin{equation}
	{\bf{S}} = \left[ {{{\bf{s}}_{{\tau _1}}},{{\bf{s}}_{{\tau _2}}}, \ldots ,{{\bf{s}}_{{\tau_{N_r}}}}} \right]^{\rm T},
\end{equation}
${\bf{y}}_n=[y_n[0],y_n[1],...,y_n[N_p-1]$ is the received time discrete signal in $n$-th antenna, ${\bf{M}} \in {{\Bbb C}^{N_r \times N_r }}$ is the mixed matrix, ${\bf{S}} \in {{\Bbb C}^{N_r \times N_p}}$ is the source matrix, ${{{\bf{s}}_{{\tau_{N_r}}}}} \in {{\Bbb C}^{1 \times N_p}}$ is the multi-path signal with different delay to be separated, and ${\bf{Z}} \in {{\Bbb C}^{N_r \times N_p}}$ is the noise matrix, $N_p$ is the number of sampling points. Then, we reconstruct the source matrix from the observation matrix by using the BSS method. However, the following conditions need to be satisfied:

\subsubsection{Conditions for BBS} 

\begin{figure}[b]
	\includegraphics[width=7.5cm]{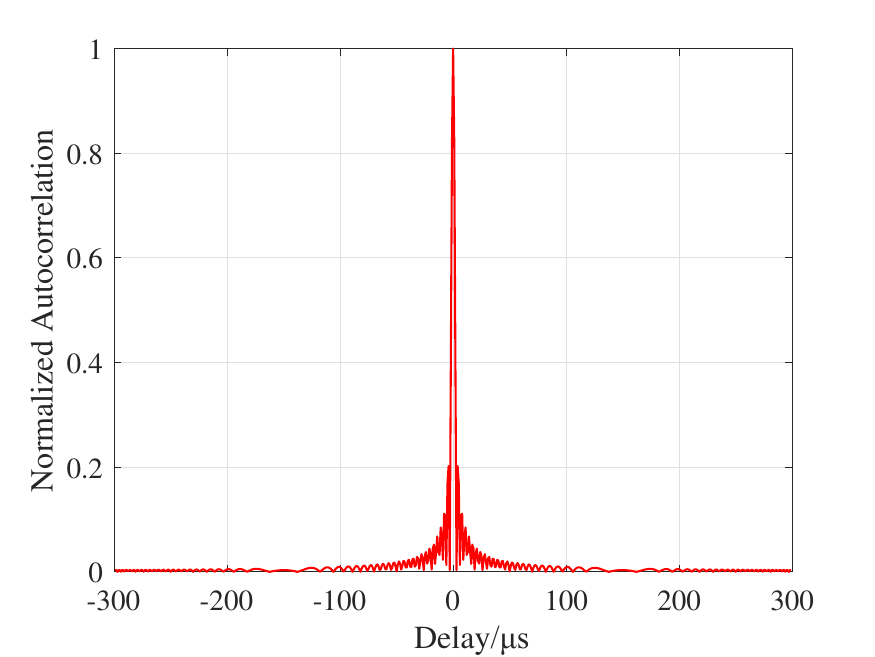}
	\caption{Normalized Autocorrelation of LFM signal.}
	\label{fig4}
\end{figure}

LFM signals of different paths and noise are regarded as the source signals and separated by joint approximative diagonalization of eigenmatrix. In order to use BBS method, the following conditions of source signals should be satisfied\cite{34}:

\textbf{a.} The source signals need to be statistically independent or uncorrelated of each other.
LFM signals of different links have different time delay, so we can determine the correlation of LFM signals with different delays by analyzing its auto-correlation function. The auto-correlation function of LFM signal can be expressed as
\begin{equation}\label{func_12}
	{r_{xx}}(\tau ) = \left| {\left( {1 - \frac{{|\tau |}}{{\tau '}}} \right){\text{sinc}}\left( {\pi k\tau \tau '\left( {1 - \frac{{|\tau |}}{{\tau '}}} \right)} \right)} \right| \ {\rm with}\ |\tau| < \tau '
\end{equation}
where $\tau $ is the time delay, $\tau '$ is the length of the LFM signal, $k$ is the chirp rate.
Figure.~\ref{fig4} is the normalized auto-correlation function of LFM signal. It can be seen that LFM signals with different delays have low correlation. Therefore, condition {\bf a} is satisfied.

\textbf{b.} There can be at most one signal following Gaussian distribution in the source signals.	
LFM signal is the sub-Gaussian signal. So only noise is the Gaussian signal. Therefore, condition {\bf b} is satisfied.

\textbf{c.} The mixed matrix ${\bf{M}}$ have to be an invertible matrix.
To satisfy condition {\bf c}, we need to specify the number of source signals. Multiple scholars have put forward methods of estimating the number of sources and have achieved acceptable results, such as spatial smoothing rank (SSR)  \cite{35}, information theory \cite{36}, singular value decomposition (SVD)  \cite{37}. The estimation of the number of source signals is not the focus of this paper, so the specific methods are not described here. The communication users takes the same number of antennas as the source signals for observation, so that the mixed matrix is reversible. Therefore, condition {\bf c} is satisfied.

\subsubsection{Multi-path LFM signals separation method}After we get the observation matrix, a method based on joint approximative diagonalization of eigen matrix is used to separate multi-path LFM signals, but signal preprocessing should be first completed.

\textbf{a.} Signal preprocessing

The observed signal is preprocessed in order to decrease redundant information and noise. The preprocessing including centralization and whitening. The observation matrix after centralization can be expressed as

\begin{equation}\label{func_13}
	{{\bf{\hat Y}}} = [{\bf{\hat y}}_1,{\bf{\hat y}}_2,...,{\bf{\hat y}}_{N_r}]^T,
\end{equation}
where
\begin{equation}\label{func_14}
	{{\bf{\hat y}}_i} = {{\bf{y}}_i} - \mathbb{E}\{ {{\bf{y}}_i}\} ,
\end{equation}
${{\mathbf{Y}}_{r}} \in {\mathbb{C}^{N_r \times N_r}}$ is the covariance matrix of ${\bf{\hat Y}}$, ${{\bf{\Phi }}_x} \in {\mathbb{C}^{N_r \times N_r}}$ is a  matrix composed of eigenvectors of ${{\bf{Y}}_r}$  and ${{\mathbf{\Sigma }}_r} \in {\mathbb{C}^{N_r \times N_r}}$ is a diagonal matrix composed of eigenvalues ${\sigma _i}$ of ${{\bf{Y}}_r}$.
Then, the whitening matrix  ${\bf{W}}$ can be calculated as 
\begin{equation}\label{func_15}
	{\bf{W}} = {{\bf{\Phi }}_r}{\bf{\Sigma }}_r^{ - 1/2}{\bf{\Phi }}_r^{\rm{H}},
\end{equation}
where
\begin{equation}\label{func_16}
	{{\bf{\Sigma }}_r} = {\rm diag}({\sigma _1},{\sigma _2}, \ldots ,{\sigma _{N_r}}).
\end{equation}
Of note, $\bf{W}$ satisfies the following condition
\begin{equation}\label{func_17}
	{\bf{W}} {\bf{M = U}},
\end{equation}
where ${\bf{U}} \in {{\Bbb C}^{{N_r} \times {N_r}}}$ is a unitary matrix. The observation matrix after preprocessing can be expressed as
\begin{equation}\label{func_18}
	{\bf{\tilde Y}} = {\bf W} {\bf{\hat Y}}.
\end{equation}
\textbf{b.} Joint approximative diagonalization of eigen matrix

In order to recover the source matrix ${\bf{S}}$ , we calculate the unitary matrix ${\bf{U}}$ in (22). It can be proved that the fourth order cumulant matrix of ${\bf{\tilde Y}}$ can be diagonalized by the unitary matrix ${\bf{U}}$, so the unitary matrix ${\bf{U}}$ can be calculated by diagonalize the set of the fourth order cumulant matrices of ${\bf{\tilde Y}}$ jointly. Fourth order cumulant function can be defined as
\begin{equation}\label{func_19}
	\begin{array}{l}
		{\rm{cum}}({x_i},x_k^*,{x_l},x_n^*) = {\rm{{\mathbb E}\{ }}{x_i}x_k^*{x_l}x_n^*{\rm{\}  - {\mathbb E}\{ }}{x_i}x_k^*\} {\rm{{\mathbb E}\{ }}{x_l}x_n^*{\rm{\} }}\\
		{\rm{ - {\mathbb E}\{ }}{x_i}{x_l}\} {\rm{{\mathbb E}\{ }}x_k^*x_n^*{\rm{\}  - {\mathbb E}\{ }}{x_i}x_n^*\} {\rm{{\mathbb E}\{ }}{x_l}x_k^*{\rm{\} }}.
	\end{array}
\end{equation}

For any $N_r \times N_r$ dimension, a nonzero matrix ${\bf{G}} = {{\rm{[}}{{g}_{{{ij}}}}{\rm{]}}_{N_r \times N_r}}$  can define the fourth order cumulant matrix ${{\bf{Q}}_{{\rm{\tilde Y}}}}({\bf{G}}) = {[{a_{i,j}}]_{N_r \times N_r}}$, where ${a_{i,j}}$ can be expressed as
\begin{equation}\label{func_20}
	{a_{ij}} = \sum\limits_{l = 1}^{N_r} {\sum\limits_{n = 1}^{N_r} {{\rm{cum}}({{\tilde r}_i},\tilde r_j^*,{{\tilde r}_l},\tilde r_n^*){g_{nl}}} } .
\end{equation}

Matrix ${{\bf{G}}_i}$, which satisfies ${{\bf{Q}}_{\rm \tilde Y}}({{\bf{G}}_i}) = \lambda {{\bf{G}}_i}$, can be regarded as the eigenmatrix of ${{\bf{Q}}_{\rm \tilde Y}}({{\bf{G}}_i})$, where $\lambda $ is the eigenvalue. ${{\bf{Q}}_{\rm \tilde Y}}({{\bf{G}}_i})$ has ${{N_r}^2}$ eigenvalues, but only $N_r$ eigenvalues are nonzero, and the $N_r$ corresponding eigenmatrices form the matrix set as follows:
\begin{equation}\label{func_21}
	\Omega  = \{ {\lambda _i},{{\bf{G}}_i}|1 \le i \le N_r\}  .
\end{equation}

In order to simplify the calculation, we could evaluate the unitary matrix ${\bf{U}}$ by jointly diagonalizing all the eigenmatrices ${{\bf{G}}_i}$ in the set ${\bf{\Omega }}$. Based on the Frobenius norm, the objective function can be obtained as
\begin{equation}\label{func_22}
	{\rm{J}}({\bf{U}}) = \sum\limits_{i = 1}^{N_r} {{\rm{off}}({\bf{U}}{{\bf{G}}_i}{{\bf{U}}^{\rm{H}}})} ,
\end{equation}
where
\begin{equation}\label{func_23}
	{\rm{off}}({\bf{\Phi }}) = ||{\bf{\Phi }} - {\rm{diag}}({\bf{\Phi }})||_{\rm{F}}^{\rm{2}} = \sum\limits_{i \ne j} {|{b_{i,j}}{|^2}}  ,
\end{equation}
with ${\rm{off}}({\bf{\Phi }})$ being the Frobenius norm for non-diagonal elements of ${\bf{\Phi }}$. The unitary matrix ${\bf{U}}$ can be obtained by solving the following optimization problem.
\begin{equation}\label{func_24}
	{\bf{\hat U}} = \arg  \mathop {\min }\limits_{\bf{U}} {\rm{J}}({\bf{U}})  ,
\end{equation}
By combining (15) and (22), and ignoring the noise ${\bf{Z}}$, we obtain the unmixed source signals as
\begin{equation}\label{func_25}
	{\bf{S}} = {{\bf{\hat U}}^{ - 1}}{\bf{WY}}  ,
\end{equation}
Every row vector of ${\bf{S}}$ is a single path of the LFM signal. The path with the shortest delay can be regarded as LOS link, defined as ${r_{{\rm{LOS}}}}(t)$, which is used for parameter estimation.
The computational complexity of the JADE algorithm primarily stems from three aspects:
1) Whitening processing of the observed signals: The main computational load involves the calculation of the covariance matrix and eigenvalue decomposition, with a complexity of ${\cal O}\left( {N_r^2{N_p} + N_r^3} \right)$.
2) Calculation of the fourth-order cumulant matrices: The complexity is ${\cal O}\left( {N_r^4{N_p}} \right)$.
3) Joint approximate diagonalization: The complexity is ${\cal O}\left( {N_r^4\varpi } \right)$, where $\varpi$ indicates the number of iterations \cite{34,38}.

\subsection{Parameter estimation of LFM signal}
LFM signal is a typical non-stationary signal. STFT can quickly analyze time-frequency characteristic and estimate parameters of LFM signal with low accuracy. In addition, LFM signal is the basis of FRFT; so, FRFT can estimate the parameters of LFM signal through global search, while spending much time. Combining the advantages of STFT and FRFT, we use an approach that employs STFT to estimate the parameters of LFM signal roughly to narrow the search range and FRFT is used for accurate estimation.
\subsubsection{Rough estimation by STFT}
STFT divides a long signal into several short-time signals by sliding window, and approximately considers that the signal is stationary in a time window. Fourier transform is done in each time window, which can be expressed as
\begin{equation}\label{func_26}
	{\mathcal{S}}(f,t) = \int_{ - \infty }^{ + \infty } {{r_{{\rm{LOS}}}}(\tau )w(\tau  - t)}  \cdot {{\rm{e}}^{ - j{\rm{2\pi }}f\tau }}{\rm{d}}\tau   ,
\end{equation}
where $w(t)$ is the window function. Comparing the spectrum characteristics in different time windows, the time-frequency characteristic of the signal can be extracted.  The center time corresponding to each window is ${\tau _i}$, and the maximum value of the frequency in each window is ${f_i}$, the point set is defined by $\left\{ {{\tau _i},{f_i}} \right\}$. The relationship between time and frequency of LFM signal is a linear function, so the least square method can be used to fit the point set $\left\{ {{\tau _i},{f_i}} \right\}$, i.e.
\begin{equation}\label{func_27}
	\{ {a_0},{a_1}\}  = \mathop {\arg }\limits_{{a_0},{a_1}} {\rm{   min}}\left( {\sum\limits_{i = 0}^{m - 1} {({a_0} + {a_1}{x_i} - {y_i}} {)^2}} \right)   ,
\end{equation}
By solving (32), the rough parameters estimation of the LFM signal can be obtained as follows:
\begin{equation}\label{func_28}
	\left\{ \begin{array}{l}
		{f_{{\rm{STFT}}}} = {a_0}\\
		{k_{{\rm{STFT}}}} = 2{a_1}
	\end{array} \right.   ,
\end{equation}
Although the STFT is limited, it provides the initial parameters for the fine estimation and narrow the search range. 
\subsubsection{Fine estimation by FRFT}
The LFM signal is the basis of FRFT, the signal to be measured will show the best focusing performance after FRFT, if the chirp rate is the same as the basis function of FRFT.
The $p$-th order FRFT of can be defined as in\cite{39}
\begin{equation}\label{func_29}
	{x_p}(u) = \int_{ - \infty }^{ + \infty } {{r_{{\rm{LOS}}}}(t)  {K_p}(t,u){\rm{d}}t}   ,
\end{equation}
where
\begin{equation}\label{func_30}
	{K_p}(t,u) = \left\{ {\begin{array}{*{20}{l}}
			{{{\text{e}}^{{\text{j}}\frac{{\cot \alpha \left( {{t^2} + {u^2}} \right)}}{2} - ut\sin \alpha }},{\text{for }}\alpha  \ne n\pi } \\ 
			{\delta \left( {t - u} \right),{\text{for }}\alpha  = 2n\pi } \\ 
			{\delta \left( {t + u} \right),{\text{for }}\alpha  = \left( {2n \pm 1} \right)\pi } 
	\end{array}} \right.,
\end{equation}
${K_p}(t,u)$ is the core function, while
\begin{equation}\label{func_31}
	{A_\alpha } = \sqrt {\frac{{(1 - {\text{j}}\cot \alpha )}}{{2\pi }}}  ,
\end{equation}
and
\begin{equation}\label{func_32}
	\alpha  = \frac{{p\pi }}{2} ,
\end{equation}
The peak point of ${r_{{\rm{LOS}}}}(t)$ in the FRFT transform domain is denoted as $(u,|{X_\alpha }(u){|^2})$. If  $\alpha  =  - {\mathop{\rm arccot}\nolimits} k$, the peak value of ${r_{{\rm{LOS}}}}(t)$ in FRFT transform domain is the maximum value, as shown in Fig.~\ref{fig5}. ${F_\alpha }$ is defined as the maximum value of ${r_{{\rm{LOS}}}}(t)$ in the FRFT transform domain when the rotation angle is $\alpha $. Bisection method is used to solve the following optimization problem:
\begin{figure}[t]
	\includegraphics[width=8.5cm]{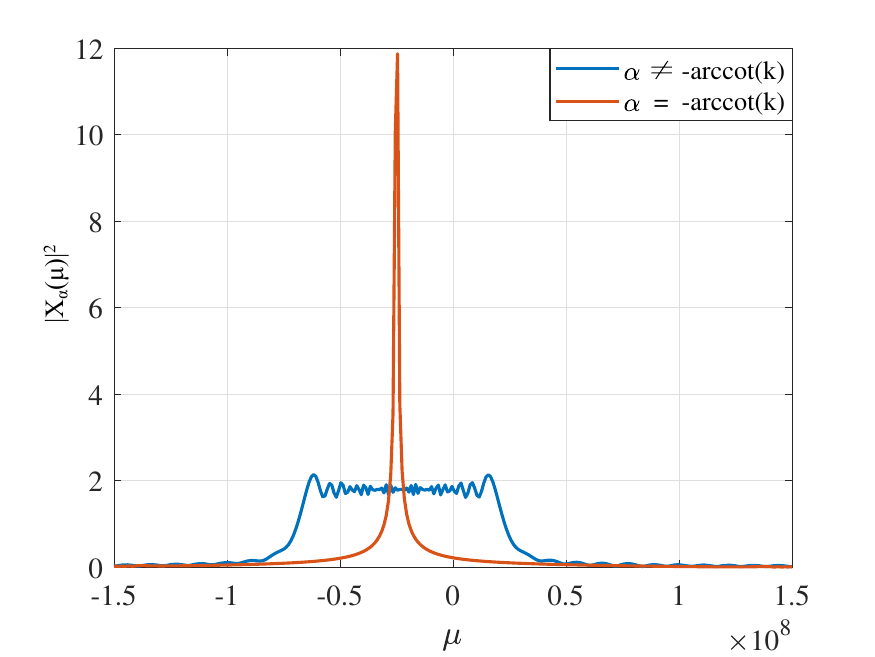}
	\caption{FRFT of LFM signal in different order.}
	\label{fig5}
\end{figure}

\begin{equation}\label{func_33}
	\left\{ {\hat \alpha ,\hat u} \right\} = \arg  \mathop {\max }\limits_{\alpha ,u} |{X_\alpha }(u){|^2} ,
\end{equation}
and the rough parameters estimation of ${r_{{\rm{LOS}}}}(t)$ is used to be the start point of bisection method as follows:
\begin{equation}\label{func_34}
	{\alpha _{{\rm{mid}}}} =  - {\mathop{\rm arccot}\nolimits} ({k_{{\rm{STFT}}}}).
\end{equation}

Then the estimation of chirp rate, initial frequency and amplitude for ${r_{{\rm{LOS}}}}(t)$ can be obtained 
\begin{equation}\label{func_35}
	{k_{{\rm{FRFT}}}} =  - \cot \hat \alpha ,
\end{equation}
\begin{equation}\label{func_36}
	{f_{{\rm{FRFT}}}} = \hat u\csc \hat \alpha  ,
\end{equation}
\begin{equation}\label{func_37}
	{A_{{\rm{FRFT}}}} = \frac{{|{X_{\hat \alpha }}(\hat u)|}}{{T|{A_{\hat \alpha }}|}} ,
\end{equation}
\begin{equation}\label{func_38}
	{r_{{\rm{FRFT}}}}(t) = {A_{\rm FRFT}}\exp \left( {{\rm j}2{\rm{\pi }}\left( {{f_{{\rm{FRFT}}}}t + \frac{1}{2}{k_{{\rm{FRFT}}}}{t^2}} \right)} \right) .
\end{equation}
\subsubsection{Complexity analysis} 
The rough parameter estimation of the LFM signal by STFT with ${N_p}$ sampling points, $({N_p} - {N_w})/{N_c} + 1$  DFT operations are required, where ${N_w}$ is the sampling point number of the window, ${N_c}$ is the sampling point number of the step. Then, the complexity of STFT is ${{\cal O}}\left( {\left( {({N_p} - {N_w})/{N_c} + 1} \right)  {N_w}\log {N_w}} \right)$.
The fine parameter estimation of the LFM signal by FRFT with ${N_p}$ sampling points, $2{\log _2}\left( {\lambda /\varepsilon } \right)$ FRFT operations are required. Then, the complexity of FRFT is ${{\cal O}}\left( {2  {{\log }_2}\left( {\lambda /\varepsilon } \right)  {N_p}  \log ({N_p})} \right)$.
\subsection{Revision of estimated LFM parameters}
FRFT accurately estimates the parameters of LFM signal, but there is still a non-negligible deviation in the estimation result due to the quantization error and noise. Moreover, the deviation accumulates over time. In particular, the LFM parameter estimation error causes a large channel estimation error. Therefore, a parameter revision method is proposed based on  the statistical characteristics of the LFM signal.

Kurtosis is defined as the fourth central moment of a random variable divided by the fourth power of the standard deviation, which is used to test the deviation of the signal from the normal distribution.
Kurtosis can be expressed as
\begin{equation}\label{func_39}
	{K_4}(x) = \frac{{\frac{1}{n}\sum\limits_{i = 1}^n {{{\left( {{x_i} - \mathbb{E}(x)} \right)}^4}} }}{{{{\left( {\frac{1}{n}\sum\limits_{i = 1}^n {{{\left( {{x_i} - \mathbb{E}(x)} \right)}^2}} } \right)}^2}}} .
\end{equation}
The kurtosis value of the normal distribution is 3. The LFM signal is a sub-Gaussian signal, and its kurtosis value is less than 3. 
${r_{{\rm{LOS}}}}(t)$ is composed of sub-Gaussian LFM signal and Gaussian equivalent noise (OFDM signal and Gaussian white noise). If the LFM signal parameters are accurately estimated and removed from ${r_{{\rm{LOS}}}}(t)$, the remaining signal follows  Gaussian distribution with a kurtosis of 3. Thus, the objective function is constructed as 
\begin{equation}\label{func_40}
	\Gamma  = {\left( {{K_4}({r_{{\rm{LOS}}}}(t) - {r_{{\rm{FRFT}}}}(t)) - 3} \right)^2} ,
\end{equation}
SA algorithm is used to solve the following optimization problem:
\begin{equation}\label{func_41}
	\left\{ {\hat A,\hat f,\hat k} \right\} = \arg  \mathop {\min }\limits_{A,f,k} {\rm{ }}\Gamma  ,
\end{equation}
where $\hat A,\hat f,\hat k$ are the final estimation result of initial frequency and chirp rate. The estimated LFM signal can be expressed as 
\begin{equation}\label{func_42}
	\hat s(t) = {\hat A}\exp \left( {2j{\rm{\pi }}\left( {\hat ft + \frac{1}{2}\hat k{t^2}} \right)} \right)  ,
\end{equation}
Of note, there exists the amplitude uncertainty of BSS. It does not affect the time-frequency characteristics of LFM signals.

\section{sensing-assisted channel estimation}
In this section, a sensing-assisted channel estimation method based on cyclic maximum likelihood is studied. The sensing signal at the BS can be accurately reconstructed after parameter estimation. Therefore, the sensing signal is considered as known, used for channel estimation as superimposed pilot.

In  conventional channel estimation methods, the pilot and the transmission signal are separated in time domain or in the frequency domain \cite{40}. However, the sensing signal and the communication signal are superimposed in both time and frequency domain. Thus, we need to consider the interference of the communication signal on the channel estimation. Although this increases the complexity of channel estimation, it can effectively improve the utilization of time-frequency resources.
\subsection{Channel estimation based on CML}
According to the central limit theorem, OFDM signal is regarded as noise, and the discrete received signal in time domain can be written as
\begin{equation}\label{func_44}
	{\bf{y}} = {\bf{S}}_0  {\bf{h}} + {\bf{z}}  ,
\end{equation}
where
\begin{equation}\label{func_45}
{\bf{y}} = {[y[0],y[1], \ldots ,y[N_p - 1]]^{\rm{T}}}  ,
\end{equation}
\begin{equation}\label{func_46}
	{\mathbf{S_0}} = {\left[ {\begin{array}{*{20}{c}}
				{s[0]}&0&0& \cdots &0 \\ 
				{s[1]}&{s[0]}&0& \cdots &0 \\ 
				{s[2]}&{s[1]}&{s[0]}& \cdots &0 \\ 
				\vdots & \vdots & \vdots & \ddots & \vdots  \\ 
				{s[l - 1]}&{s[l - 2]}&{s[l - 3]}& \cdots &{s[0]} \\ 
				{s[l]}&{s[l - 1]}&{s[l - 2]}& \cdots &{s[1]} \\ 
				\vdots & \vdots & \vdots & \ddots & \vdots  \\ 
				0&0&0& \cdots &{s[{N_p} - l]} 
		\end{array}} \right]} ,
\end{equation}
\begin{equation}\label{func_47}
	{\bf{h}} = {[h[0],h[1], \ldots ,h[l - 1]]^{\rm{T}}}  ,
\end{equation}
\begin{equation}\label{func_48}
	{\bf{z}} = {[z[0],z[1], \ldots ,z[N_p - 2],z[N_p - 1]]^{\rm{T}}}.
\end{equation}
In (49)-(52) $\left\{ {s[n]} \right\}_{n = 0}^{{N_p} - 1}$ is the discrete LFM signal,  $l$ is the length of channel delay,  ${\bf{z}}$ follows complex Gaussian distribution with a mean of 0 and a variance of $\sigma _z^2 $.  ${{\bf{y}}\in {{\Bbb C}^{N_p \times 1}}}$ follows complex Gaussian distribution with a mean of ${\bf{\mu }} = {\bf{Sh}}$ and a covariance of ${\bf{R=\mathbb{E}[(y-Sh)(y-Sh)^{\rm H}]}}$. The probability density function of ${\bf{y}}$ can be expressed as
\begin{equation}\label{func_49}
f({\bf{y;\mu ,R}}) = \frac{{\exp \left( { - \frac{1}{2}({\bf{y}} - {\bf{\mu }}){{\bf{R}}^{ - {\bf{1}}}}{{({\bf{y}} - {\bf{\mu }})}^{\rm{H}}}} \right)}}{{{{(2{\rm{\pi }})}^{M/2}}{{\bf{R}}^{1/2}}}}  .
\end{equation}
Let $M$ be the number of samples, the likelihood function of $\bf{h} $ can be obtained as
\begin{equation}\label{func_50}
p({\bf{y}}|{\bf{h,R}}) = \prod\limits_{m = 1}^M {f({{\bf{y}}_m}|{\bf{h,R}})}   .
\end{equation}
The Log-likelihood function obtained as
\begin{equation}\label{func_51}	
	\begin{split}
	L({\bf{h,R}}) &=  - \ln \left( {p({\bf{y}}|\bf{h,R})} \right) \\
	&=  - \sum\limits_{m = 1}^M {\ln \left( {f({{\bf{y}}_{m}}|\bf{h,R} )} \right)}   .
	\end{split}
\end{equation}

Since the covariance matrix $\bf R$ is coupled with $\bf h$, it is hard to directly obtain the maximum likelihood estimate of $\bf h$. Therefore, we adopt an alternating optimization approach that follows the following steps:

{\bf Step 1}: Initialize $\bf R=I$, and  obtain
\begin{equation}\label{func_52}
	 L({\bf{h}}) \propto  - \frac{1}{2}\sum\limits_{m = 1}^M {\left( {{{\bf{y}}_m} - {\bf{S_0h}}} \right){{\bf{R}}^{ - 1}}{{\left( {{{\bf{y}}_m} - {\bf{S_0h}}} \right)}^{\text{H}}}} .
\end{equation}

{\bf Step 2}: Set the partial derivative of (56) to be equal to $0$, i.e., 
\begin{equation}
	\frac{\partial }{{\partial {\bf{h}}}}\left\{ { - \frac{1}{2}\sum\limits_{m = 1}^M {\left( {{{\bf{y}}_m} - {\bf{S_0h}}} \right){{\bf{R}}^{ - 1}}{{\left( {{{\bf{y}}_m} - {\bf{S_0h}}} \right)}^{\text{H}}}} } \right\} = 0,
\end{equation}
and obtain the maximum likelihood estimation of $\mathbf{h}$ as
\begin{equation}
	{\bf{\hat h}} = {\left( {{{\bf{S_0}}^{\text{H}}}{{\bf{R}}^{ - 1}}{\bf{S_0}}} \right)^{ - 1}}{{\bf{S_0}}^{\text{H}}}{{\bf{R}}^{ - 1}}\left( {\frac{1}{M}\sum\limits_{m = 1}^M {{{\bf{y}}_m}} } \right) .
\end{equation}

{\bf Step 3}: Set ${\bf h={\hat h}}$, and obtain
\begin{equation}
\begin{aligned}
	L({\bf{R}}) \propto & -\frac{1}{2} \left( M \ln \left| {\bf{R}} \right| \right. \\
	& \left. + \sum\limits_{m = 1}^M \left( {{{\bf{y}}_m} - {\bf{S_0h}}} \right) {{\bf{R}}^{ - 1}} {{\left( {{{\bf{y}}_m} - {\bf{S_0h}}} \right)}^{\text{H}}} \right).
\end{aligned}
\end{equation}

{\bf Step 4}: The partial derivative of (59) to be equal to $0$, i.e. 
\begin{equation}
	\begin{gathered}
		\frac{\partial }{{\partial {\mathbf{R}}}}\left\{ { - \frac{1}{2}M\ln \left| {\mathbf{R}} \right| + \sum\limits_{m = 1}^M {\left[ {\left( {{{\mathbf{y}}_m} - {{\mathbf{S}}_{\mathbf{0}}}{\mathbf{h}}} \right)} \right.} } \right. \hfill \\
		\left. {\left. {{{\mathbf{R}}^{ - 1}}{{\left( {{{\mathbf{y}}_m} - {{\mathbf{S}}_{\mathbf{0}}}{\mathbf{h}}} \right)}^{\text{H}}}} \right]} \right\} = 0, \hfill \\ 
	\end{gathered} 
\end{equation}
then, the maximum likelihood estimation of $\mathbf{R}$ can be obtained as
\begin{equation}
	{\bf{\hat R}} = \frac{1}{M}\sum\limits_{m = 1}^M {\left( {{{\bf{y}}_m} - {\bf{S_0h}}} \right){{\left( {{{\bf{y}}_m} - {\bf{S_0h}}} \right)}^{\text{H}}}} .
\end{equation}

{\bf Step 5}: Repeat {\bf Steps 2-4} until convergence or reaching the number of iterations. For clarity, we summarize the method in {\bf Algorithm 1}.

Further, we calculate the CRLB of channel estimation ${\bf{\hat h}}$. The Fisher information of the ${\bf{h}}$  can be expressed as
\begin{equation}
	\begin{aligned}
		J(\mathbf{h}) &= - \mathbb{E}\left[ \frac{\partial^2 L(\mathbf{h})}{\partial \mathbf{h}^2} \right] \\
		&= - \mathbb{E}\left[ \frac{\partial}{\partial \mathbf{h}} \left\{ \mathbf{S}_0^\text{H} \mathbf{R}^{-1} \sum_{m = 1}^M \left( \mathbf{y}_m - \mathbf{S}_0 \mathbf{h} \right) \right\} \right] \\
		&= M \mathbf{S}_0^\text{H} \mathbf{R}^{-1} \mathbf{S}_0.
	\end{aligned}
\end{equation}
CRLB of ${\bf{h}}$  can be obtained as
\begin{equation}\label{func_56}
	\begin{split}
		{\rm{CRLB}}({\bf{h}}) &={\rm{ trace}}\left( {J{{({\bf{h}})}^{ - 1}}} \right)\\
		&= \frac{1}{M}{\rm{trace}}{({{\bf{S}}_0^{\rm H}}  {{\bf{R}}^{ - 1}}  {\bf{S}}_0)^{ - 1}}.
	\end{split}
\end{equation}
\subsection{Convergence analysis}
 When ${\mathbf{R}} = {{\mathbf{R}}^{(r)}}$  is fixed, the subproblem becomes 
\begin{equation}
	{{\mathbf{h}}^{(r + 1)}} = \arg \mathop {\min }\limits_{\mathbf{h}} {\text{ }}L({\mathbf{h}},{{\mathbf{R}}^{(r)}}).
\end{equation}
This is a strictly convex quadratic programming problem with respect to ${\mathbf{h}}$. The global optimal solution is given by (58). This step ensures $L({{\mathbf{h}}^{(r + 1)}},{{\mathbf{R}}^{(r)}}) \leqslant L({{\mathbf{h}}^{(r)}},{{\mathbf{R}}^{(r)}})$. That is, updating ${\mathbf{h}}$  while fixing ${\mathbf{R}}$ strictly make the objective function value non-increasing.
Similarly, when  ${\mathbf{h}} = {{\mathbf{h}}^{(r + 1)}}$ is fixed, the subproblem becomes
\begin{equation}
	{{\mathbf{R}}^{(r + 1)}} = \arg \mathop {\min }\limits_{{\mathbf{R}} \succ 0} {\text{ }}L({{\mathbf{h}}^{(r + 1)}},{\mathbf{R}}).
\end{equation}
This subproblem is convex with respect to the positive definite matrix ${\mathbf{R}}$. The global optimal solution is given by (61). This step ensures $L({{\mathbf{h}}^{(r + 1)}},{{\mathbf{R}}^{(r + 1)}}) \leqslant L({{\mathbf{h}}^{(r + 1)}},{{\mathbf{R}}^{(r)}})$. That is, updating ${\mathbf{R}}$  while fixing  ${\mathbf{h}}$ strictly make the objective function value non-increasing. In summary, we can conclude that
\begin{equation}
	\begin{split}
	L({{\bf{h}}^{(r + 1)}},{{\bf{R}}^{(r + 1)}}) \le L({{\bf{h}}^{(r + 1)}},{{\bf{R}}^{(r)}})\\
	\le L({{\bf{h}}^{(r)}},{{\bf{R}}^{(r + 1)}}).
    \end{split}
\end{equation}
Since the likelihood function is $0 < p({\mathbf{y}}|{\mathbf{h}},{\mathbf{R}}) \leqslant 1$  , we can obtain
\begin{equation}
L({\mathbf{h}},{\mathbf{R}}) =  - \ln p({\mathbf{y}}|{\mathbf{h}},{\mathbf{R}}) \geqslant 0.
\end{equation}

According to the monotone convergence theorem, the convergence of proposed method can be guaranteed. Although the non-convex nature of the optimization problem prevents guarantees of global optimality, it ensures that the algorithm will eventually converge to a stationary point\cite{41}. 

\subsection{Optimal power weighting coefficient }
In the generation of ISAC signals, allocating higher power to the communication component yields a better SINR at the receiver. However, this inevitably intensifies the data interference to the superimposed pilots during channel estimation, thereby degrading the estimation accuracy. Consequently, we optimize the weighting factor $w$ with the objective of maximizing the SINR. Since the exact expression of the channel estimation error of the proposed method is hard to obtain, we leverage the previously derived CRLB for analysis, thereby deriving a closed-form solution.

 Noise Covariance Matrix R in (53) is determined by OFDM signal and Gaussian noise which can be expressed as
\begin{equation}
	{\mathbf{R}} = (\sigma _n^2 + wP){{\mathbf{I}}_{{N_p}}}.
\end{equation}
And we have
\begin{equation}
	{\text{trace}}\left( {{{\left( {{\mathbf{S}}_0^H{{\mathbf{S}}_0}} \right)}^{ - 1}}} \right) = \frac{D}{{(1 - w)P}},
\end{equation}
where $D$ is a constant, $P$ is the transmit power, and the CRLB in (63) can be rewritten as
\begin{equation}
	{\text{CRLB}}(w) = \frac{{D\left( {\sigma _n^2 + wP} \right)}}{{(1 - w)PM}},
\end{equation}
where $\sigma _n^2$ is the power of Gaussian white noise. The received signal can be expressed as
\begin{equation}
	\begin{gathered}
		{\mathbf{y}} = {\mathbf{X}}\left( {{\mathbf{\hat h}} + {\mathbf{\tilde h}}} \right) + {\mathbf{n}} \hfill \\
		= {\mathbf{\hat h}}({\mathbf{C}}  + {{\mathbf{S}}_{\mathbf{0}}} ) + {\mathbf{X\tilde h}} + {\mathbf{n}}, \hfill \\ 
	\end{gathered}  
\end{equation}
where ${\mathbf{X}},{\mathbf{C}} \in {\mathbb{C}^{{N_p} \times l}}$ are the lower triangular Toeplitz matrix consisted of transmit signal and communication signal, respectively, like ${{\mathbf{S}}_0}$ in (50),   ${\mathbf{\hat h}}$  is the estimation of channel gain vector,  ${\mathbf{\tilde h}}$ denotes the channel estimation error with the variance of $\sigma _{\tilde h}^2 = {\text{CRLB}}$ . After eliminating the superimposed pilot we have
\begin{equation}
	\begin{gathered}
		{{\bf{y}}_c} = \underbrace {{\bf{C\hat h}}}_{{\rm{desired\ signal}}} + \underbrace {{\bf{X\tilde h}}}_{{\rm{interference}}} + \underbrace {\bf{n}}_{{\rm{noise}}}. 
	\end{gathered} 
\end{equation}
The power of the useful signal is calculated as
\begin{equation}
	{P_{sig}} = \mathbb{E}\left[ {\left\| {{\bf{C\hat h}}} \right\|_2^2} \right] = \mathbb{E}\left[ {{{{\bf{\hat h}}}^{\rm{H}}}{{\bf{C}}^{\rm{H}}}{\bf{C\hat h}}} \right].
\end{equation}
Given that the communication signals are independent and uncorrelated, we have $\mathbb{E}\left[ {{{\bf{C}}^{\rm{H}}}{\bf{C}}} \right] = l  {{\bf{I}}_{l \times l}}wP$ , ${P_{sig}} = l\left\| {{\bf{\hat h}}} \right\|_2^2wP$. And the noise power is given by ${P_{noi}} = \mathbb{E}\left[ {\left\| {\bf{n}} \right\|_2^2} \right] = l\sigma _n^2$. Next, the data interference power is expressed as:
\begin{equation}
	\begin{array}{c}
		{P_{{\mathop{\rm int}} }} = \mathbb{E}\left[ {\left\| {{\bf{X\tilde h}}} \right\|_2^2} \right] = \mathbb{E}\left[ {{{{\bf{\tilde h}}}^{\rm{H}}}{{\bf{X}}^{\rm{H}}}{\bf{X\tilde h}}} \right]\\
		= \mathbb{E}\left[ {{\rm{trace}}\left( {{\bf{X\tilde h}}{{{\bf{\tilde h}}}^{\rm{H}}}{{\bf{X}}^{\rm{H}}}} \right)} \right]\\
		= {\rm{trace}}\left( {\mathbb{E}\left[ {{{\bf{X}}^{\rm{H}}}{\bf{X}}} \right]{\bf{\tilde h}}{{{\bf{\tilde h}}}^{\rm{H}}}} \right)
	\end{array}
\end{equation}
Substituting the transmit signal expression ${\rm{X = }}{\bf{C}}{\rm{ + }}{{\bf{S}}_0}$ , where ${{\bf{S}}_0}$  is constructed using LFM signals with superior autocorrelation property, the communication signals are zero-mean, independent and uncorrelated. In addition, pilot and communication signal are mutually uncorrelated, therefore, we can approximate the covariance matrix as $\mathbb{E}\left[ {{{\bf{X}}^{\rm{H}}}{\bf{X}}} \right] = l \cdot P{{\bf{I}}_{l \times l}}$. By modeling the channel estimation error via the derived CRLB in (63), the interference power simplifies to
\begin{equation}
	{P_{{\mathop{\rm int}} }} = l \cdot P \cdot {\rm{CRLB}}.
\end{equation}

Consequently, the SINR can be expressed as
\begin{equation}
	{\text{SINR = }}\frac{{wP{{\left\| {\mathbf{{\hat h}}} \right\|}^2}}}{{P \cdot {\text{CRLB}} + \sigma _n^2}} = \frac{{wP{{\left\| {\mathbf{{\hat h}}} \right\|}^2}(1 - w)M}}{{D\left( {\sigma _n^2 + wP} \right) + (1 - w)M\sigma _n^2}}.
\end{equation}

To maximize the SINR, we formulate the following optimization problem
\begin{equation}
	\begin{gathered}
		\mathop {\max }\limits_w {\text{ SINR}}{\text{ = }}\frac{{wP{{\left\| {\mathbf{{\hat h}}} \right\|}^2}(1 - w)M}}{{D\left( {\sigma _n^2 + wP} \right) + (1 - w)M\sigma _n^2}} \hfill \\
		{\text{s}}{\text{.t}}\ \ 0<w<1. \hfill \\ 
	\end{gathered} 
\end{equation}

By simplifying the objective function, we can reformulate the original optimization problem as
\begin{equation}
	\begin{gathered}
		\mathop {\max }\limits_w f(w) = \frac{{w - {w^2}}}{{{K_1} + {K_2}w}} \hfill \\
		{\text{s}}{\text{.t}}\ \ 0<w<1, \hfill \\ 
	\end{gathered} 
\end{equation}
where ${K_1} = \sigma _n^2(D + M)$ and ${K_2} = DP - M\sigma _n^2$, And the optimal  $w^*$ can be calculated as
\begin{equation}
	\begin{gathered}
		{w^*} = \frac{{\sqrt {{K_1}({K_1} + {K_2})}  - {K_1}}}{{{K_2}}} \\ 
		= \frac{{{\sigma _n}\sqrt {D(D + M)(\sigma _n^2 + P)}  - \sigma _n^2(D + M)}}{{DP - M\sigma _n^2}} \\ 
	\end{gathered} 
\end{equation}

\begin{algorithm}[t]
	\small 
	\SetAlgoLined
	\caption{Cyclic Maximum likelihood channel estimation method. }
	\textit{Input}:  $\varepsilon, {N_{it}}$    \\
    \textit{Output}: $\bf \hat h^{(r)}$ \\
	\textit{Initialization}: $\bf R=I$\\
    \While{$\left( {\left| {{{\bf{h}}^{(r)}} - {{\bf{h}}^{(r - 1)}}} \right| < \varepsilon } \right)\& \& \left( {r < {N_{it}}} \right)$ }{
		Calculate ${{\bf{\hat h}}^{(r)}}$ from (58)\\
		$\bf h=\hat h^{(r)}$ \\
		Calculate ${{\bf{\hat R}}}^{(r)}$ from (61) \\
		$\bf R=\hat R^{(r)}$ \\
		$r \leftarrow r+1$\\	
	}

\end{algorithm}

\section{Numerical Results}

In this section, the performance of sensing signal parameter estimation, sensing-assisted channel estimation and target sensing are simulated. The simulation parameters are summarized in Table 1.

\begin{table}[b]
	\centering
	\caption{Parameters of system}
	\setlength{\arrayrulewidth}{0.4mm}
	\setlength{\tabcolsep}{16pt}
	\renewcommand{\arraystretch}{1.5}
	\begin{tabular}{c c c c c } 
		\hline
		 Parameter & Value  \\ [0.5ex] 
		\hline
		Number of subcarriers ($N_c$) & 256    \\ 
		Number of sampling point ($N_p$)  & 400\\
		Bandwidth ($B$) & 20 MHz \\
		Pulse width of LFM ($T_s$)  & 2 $\mu$s \\
		
		Initial frequency of LFM  ($f_0$)  & 80 MHz     \\
		
		Chirp rate of LFM ($k$) & $10 \ {\rm{MHz/ \mu s}}$  \\
		\hline
	\end{tabular}
	
	\label{tab1}
\end{table}

\subsection{Parameter estimation of LFM signal }
\subsubsection{Separation of multi-path ISAC signals}
It is assumed that the ISAC signal is transmitted from  the BS to the user through 3 paths with the delays of 0, 10 and 30 sampling points, respectively. At the BS, 3 antennas are used to receive the multi-path ISAC signals. 

\begin{figure}[hb]
	\includegraphics[width=10cm]{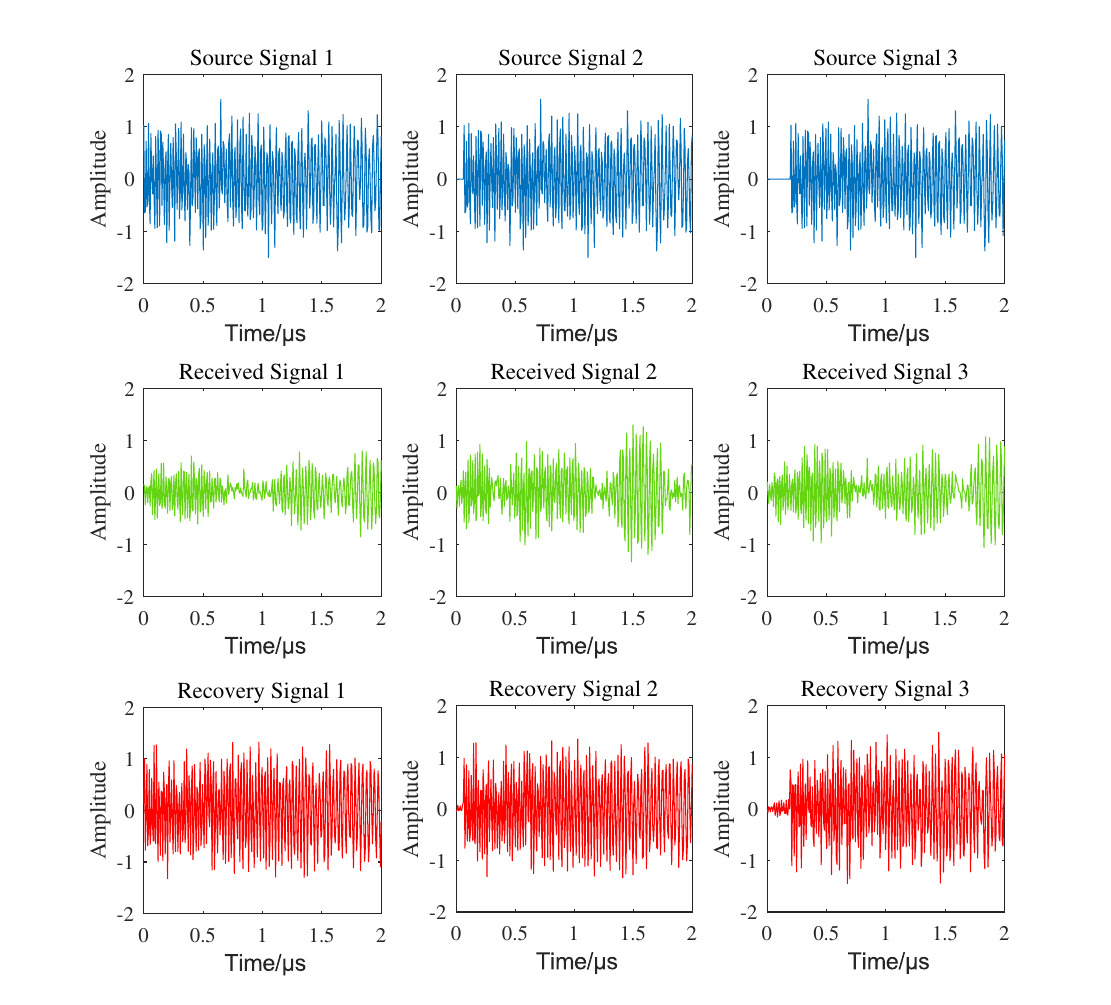}
	\caption{Separation of the multi-path ISAC signals.}
	\label{fig6}
\end{figure}

Figure 6 depicts the separation of the multi-path ISAC signals. The power weighting coefficient is 0.2 and SNR is set to $20\,\rm{dB}$. The first line are the source signals with different paths and delays, which are formed by the superposition of LFM signals, OFDM signals and Gaussian white noise. The second line are the signals received by different antennas. The third line is the recovery signals. From this figure, it becomes evident that the proposed method can effectively separate the original signals in different links with different delays.
\begin{figure}[t]
	\includegraphics[width=8.5cm]{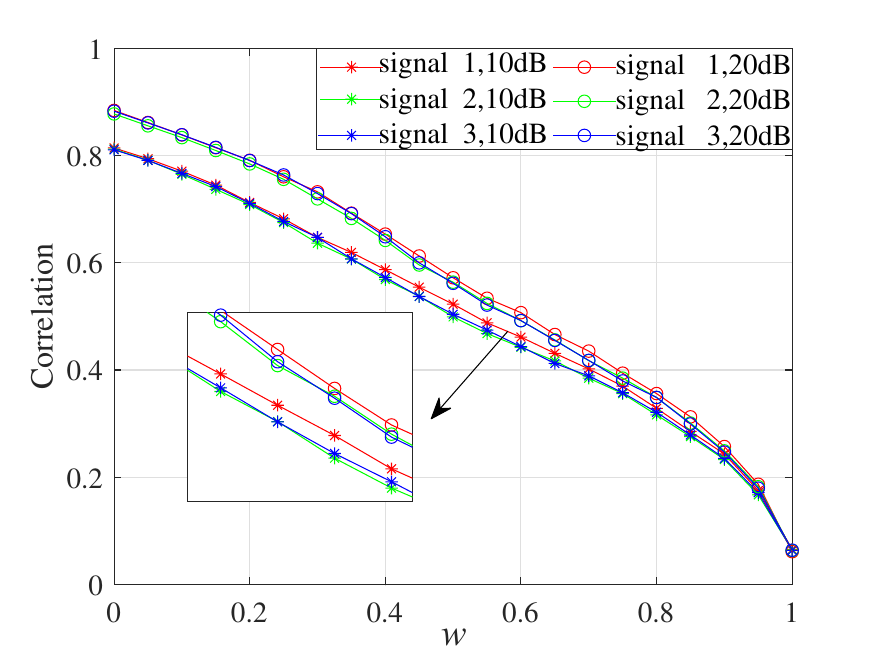}
	\caption{Correlation of source signals and recovery signals.}
	\label{fig7}
\end{figure}

In Fig. 7, the correlation between source and recovery signals as a function of the power weights for different values of SNR are presented. The correlation between the recovery signals and the source signals is used to measure the performance of signal separation with different SNR value and different power weighting coefficients. It can be seen that the signal separation performance improves as the SNR increases. On the other hand, as the power weights increase a performance degradation in terms of correlation is observed.
It is because that the source signal becomes more like a Gaussian signal with the power of noise and communication signal increasing, so that the separation performance becomes worse according to  condition b in Section \textrm{III}. When $w$=1, there is no sensing signal component in the source signal, and the source signals are all Gaussian signals; as a result, no separation is performed.
\subsubsection{Parameter estimation of LFM signal by STFT-FRFT}
\begin{figure}[t]
	\includegraphics[width=8.5cm]{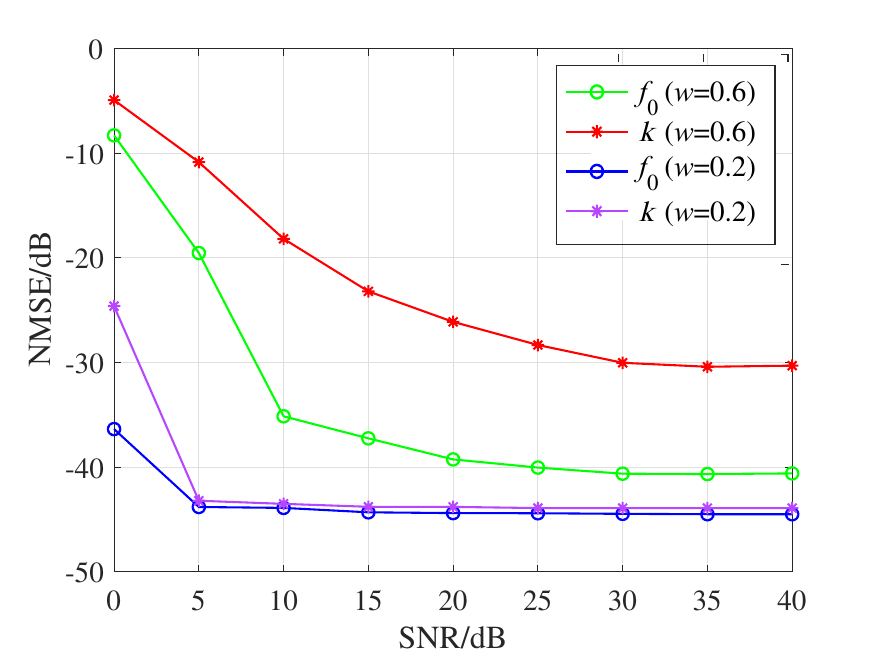}
	\caption{Parameter estimation error of STFT-FRFT.}
	\label{fig8}
\end{figure}

The separated signal with the minimum delay is regarded as the LOS signal. 
STFT-FRFT is used for parameter estimation with different SNR where $w = 0.2$ and $w = 0.6$ respectively. And the simulations are shown in Fig.~\ref{fig8}. The  NMSE of parameter estimation decreases as the SNR increases. The accuracy of parameter estimation reaches its maximum value for $w = 0.2$ and SNR  greater than 5 dB, as well as  $w = 0.6$ and SNR  greater than 30 dB. This is because the ISAC signal contains fixed communication signal, which is regarded as noise when estimating the LFM parameters. Meanwhile, there is a quantization error in FRFT, so the parameter estimation accuracy is upper bounded. 

\subsubsection{Revision of estimated LFM parameters}
The estimated LFM parameters is revised using SA. Moreover, the parameter estimation performance of the method proposed in this paper is compared with the CCT algorithm proposed in \cite{32}. Simulations are presented in Fig. 9, where the NMSE is plotted against SNR for different values of $w$.
\begin{figure}[h]
	\centering
	\subfigure[$w = 0.2$]{
		\begin{minipage}[b]{0.5\textwidth}
			\includegraphics[width=8.5cm]{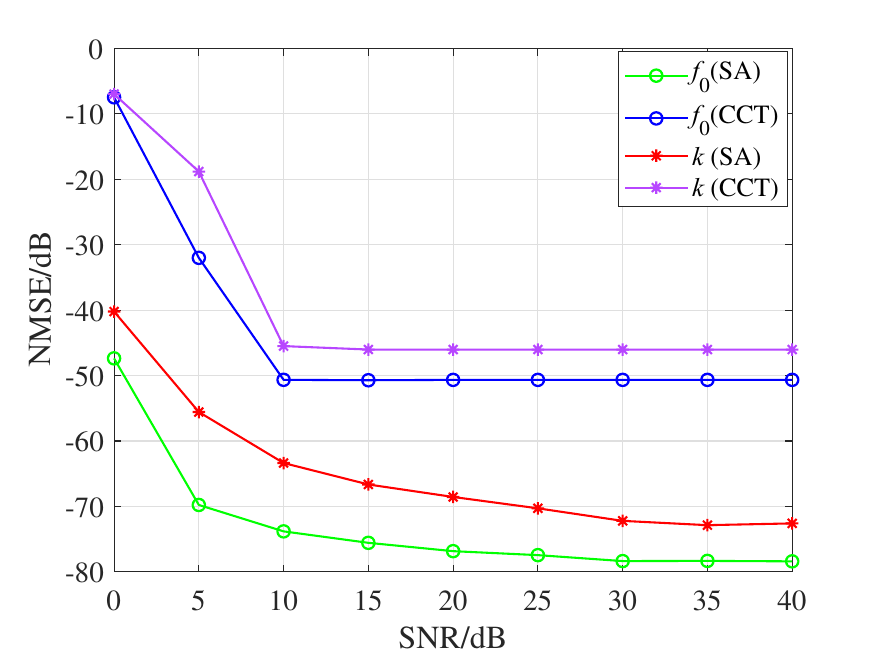}
		\end{minipage}
	}
	\subfigure[$w = 0.6$]{
		\begin{minipage}[b]{0.5\textwidth}
			\includegraphics[width=7.6cm]{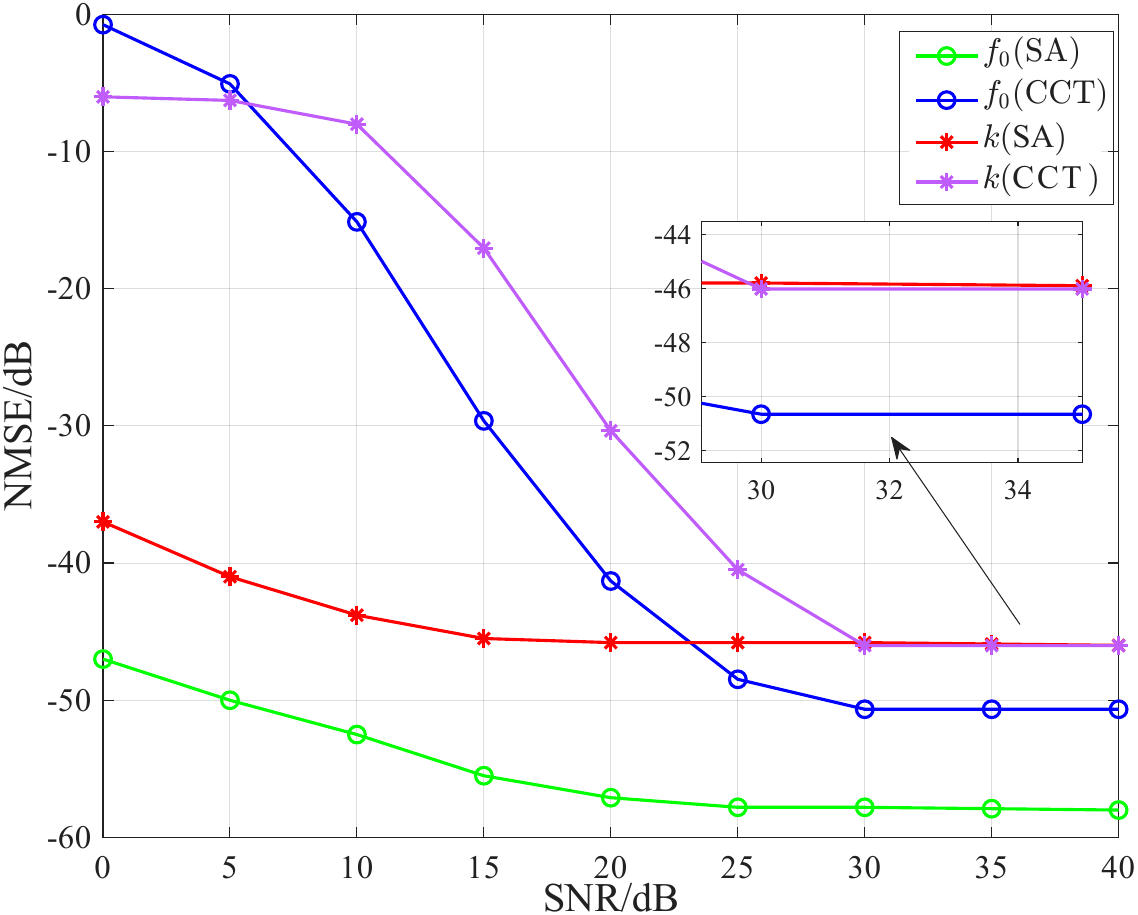}
		\end{minipage}
	}
	\caption{Revision of Parameter estimation error.}
    \label{fig9}
\end{figure}

By Comparing Fig.~\ref{fig8} and Fig.~\ref{fig9}, we observe that parameters estimation error of LFM signal by STFT-FRFT can be effectively reduced by the revision method presented in this paper.
Meanwhile, compared with the CCT method reported in \cite{32}, the NMSE of parameter estimation is greatly reduced, especially in the low SNR scenario. Since in the low SNR regime, the noise has the deeper impact on cyclic spectrum characteristic than statistical characteristic of LFM signal.
\begin{figure}[h]
	\includegraphics[width=8.5cm]{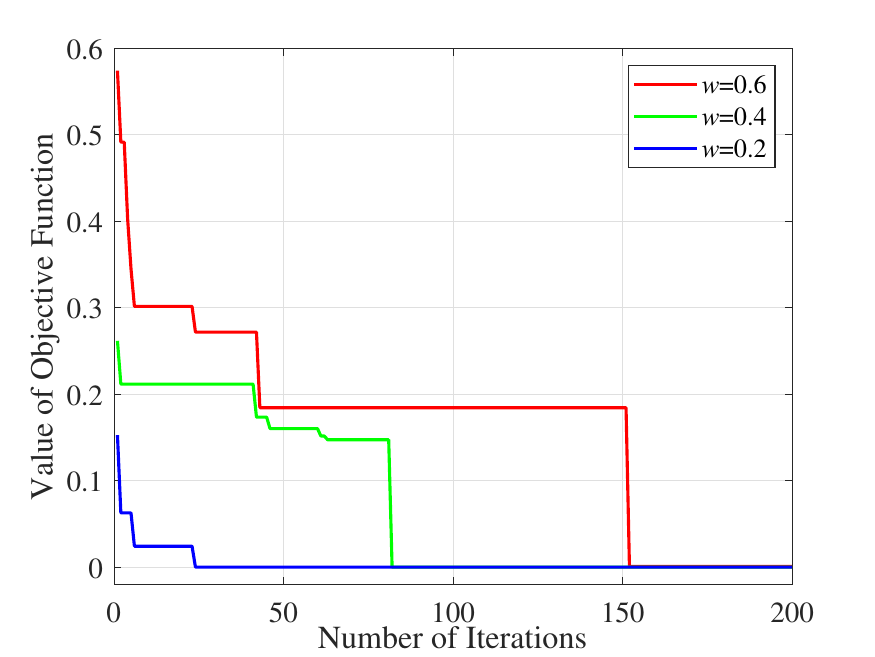}
	\caption{Convergence analysis.}
	\label{fig10}
\end{figure}

In Fig. 10, we analyze the convergence of the parameter revision algorithm. It is observed that it has the minimum initial value and the fast convergence speed for $w = 0.2$. Since the parameter estimation result by STFT-FRFT has the minimum error, for  $w = 0.2$,  according to Fig.~\ref{fig10}.  Finally, after about 150 iterations, the algorithm  converges for any power weighting coefficient.
\subsection{Analysis of communication performance}
The LFM signal is regarded as a superimposed pilot for channel estimation.  MSE of the proposed channel estimation method is compared with another classical method based on time domain least square (TDLS) for superimposed pilot  in \cite{42}. 

\begin{figure}[h]
	\includegraphics[width=8.5cm]{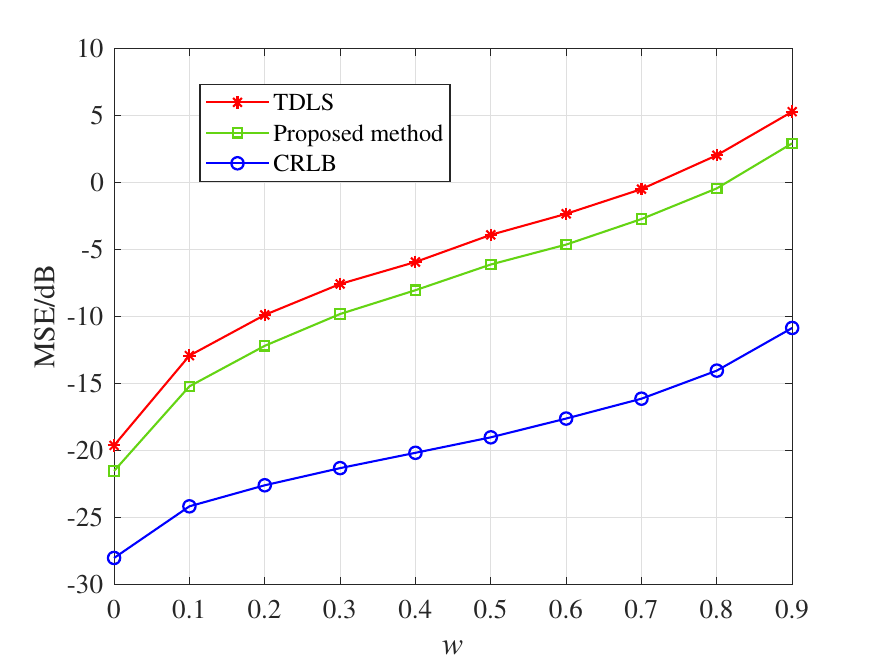}
	\caption{MSE of channel estimation with different $w$'s.}
	\label{fig11}
\end{figure}
Channel estimation performance is simulated under different power weighting coefficient, which is shown in Fig.~\ref{fig11}. With the increase of $w$, the sensing signal power for channel estimation decreases, and lead to the deterioration of channel estimation performance. Meanwhile, the MSE of channel estimation of the proposed method is about 2 dB lower than that of TDLS method, which is closer to CRLB.

\begin{figure}[htpb]
	\includegraphics[width=8.5cm]{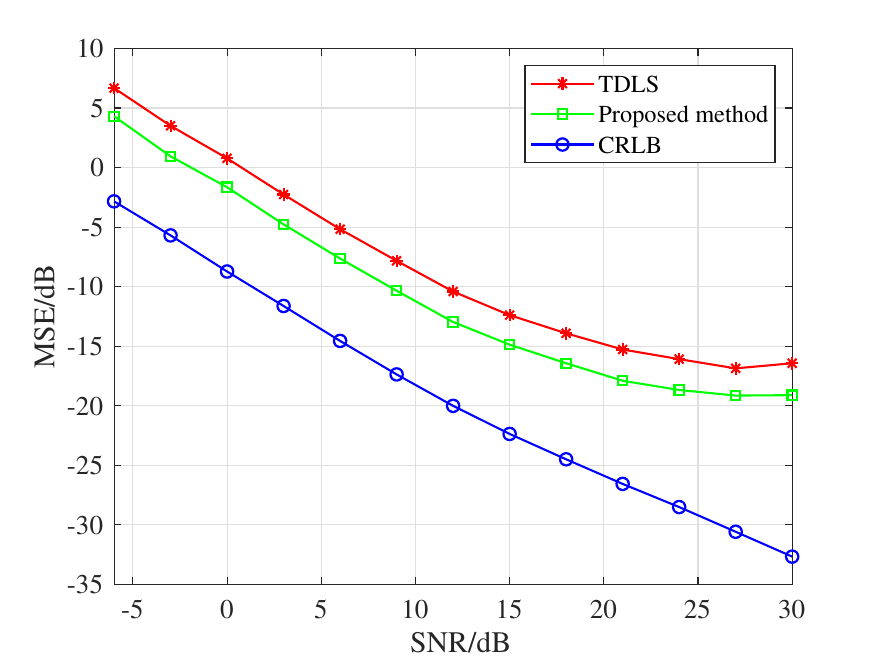}
	\caption{MSE of channel estimation with different SNR's.}
	\label{fig12}
\end{figure}

Then, we simulate the channel estimation performance for $w=0.2$ and different SNR values. The MSE of channel estimation is shown in Fig.~\ref{fig12}. From this figure. it can be observed that the channel estimation  MSE  is monotone decreasing with the increase of SNR. When the SNR is high, the channel estimation performance reaches a plateau and shows no significant further improvement. This phenomenon can be attributed to two main factors: first, the OFDM component in the ISAC signal acts as noise during the channel estimation, which limits the estimation accuracy; second, error accumulation occurs during the reconstruction of the LFM signal.
Meanwhile, the MSE  of the proposed method is about 3 dB lower than that of TDLS method, which is closer to CRLB.

\begin{figure}[htpb]
\includegraphics[width=7.5cm]{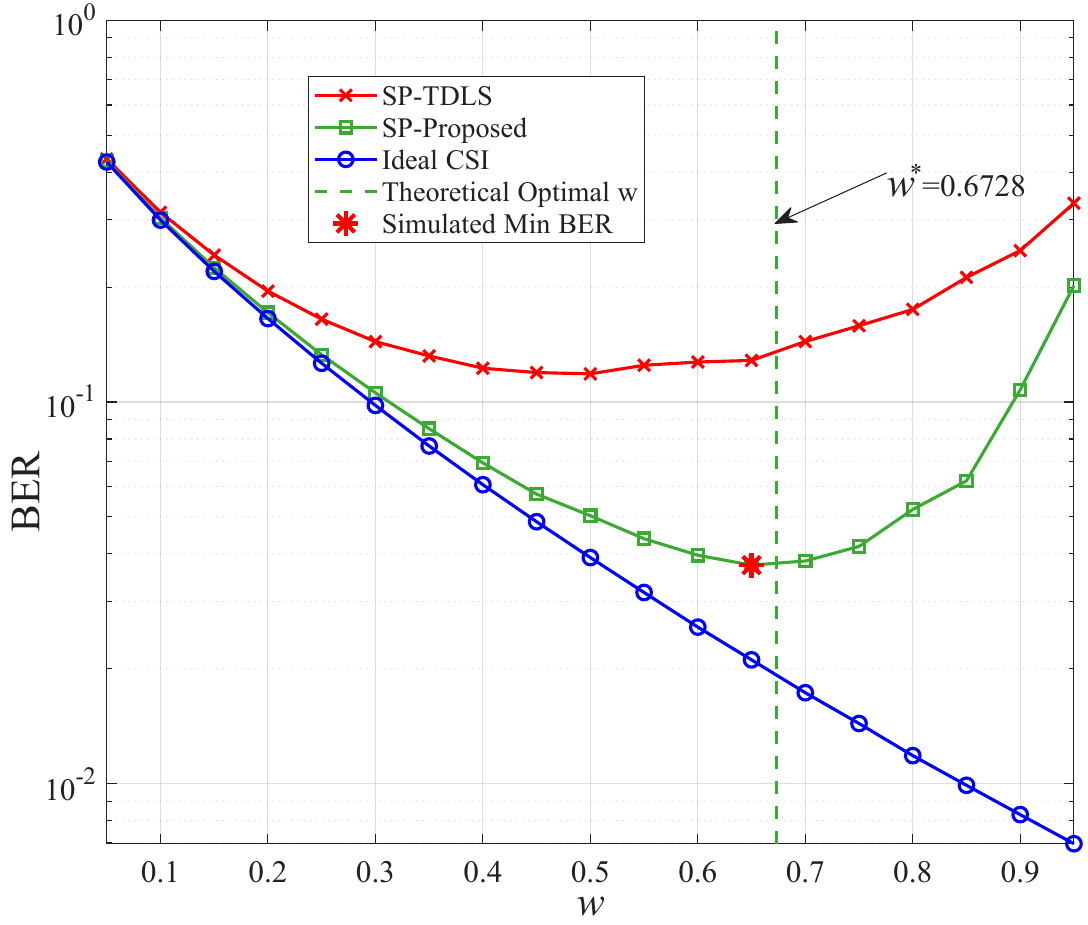}
	\caption{ BER performance with different $w$'s.}
	\label{fig13}
\end{figure}

In Fig. 13, the BER is plotted against the power weight coefficient for different estimation methods assuming that the SNR is equal to 15 dB. Here, "Ideal" refers to the performance upper bound achieved when perfect CSI is available.  It is observed that the proposed method outperforms TDLS, and is close to ideal channel estimation. For $w=0$, the power of communication signal is 0, and no communication information is transmitted.    Since the BER is a monotonically decreasing function of the SINR, maximizing the SINR is equivalent to minimizing the BER. It can be observed that the theoretical optimum $w^*$ is located close to the minimum point of the BER curve, which validates the accuracy of our estimation. It is worth noting that a slight deviation exists between the theoretical and simulated values. This discrepancy stems from the fact that our theoretical derivation utilizes the CRLB of the channel estimation error rather than the exact error variance. Nevertheless, the simulation results indicate that this deviation is within an acceptable range.

\begin{figure}[htpb]
\includegraphics[width=7.5cm]{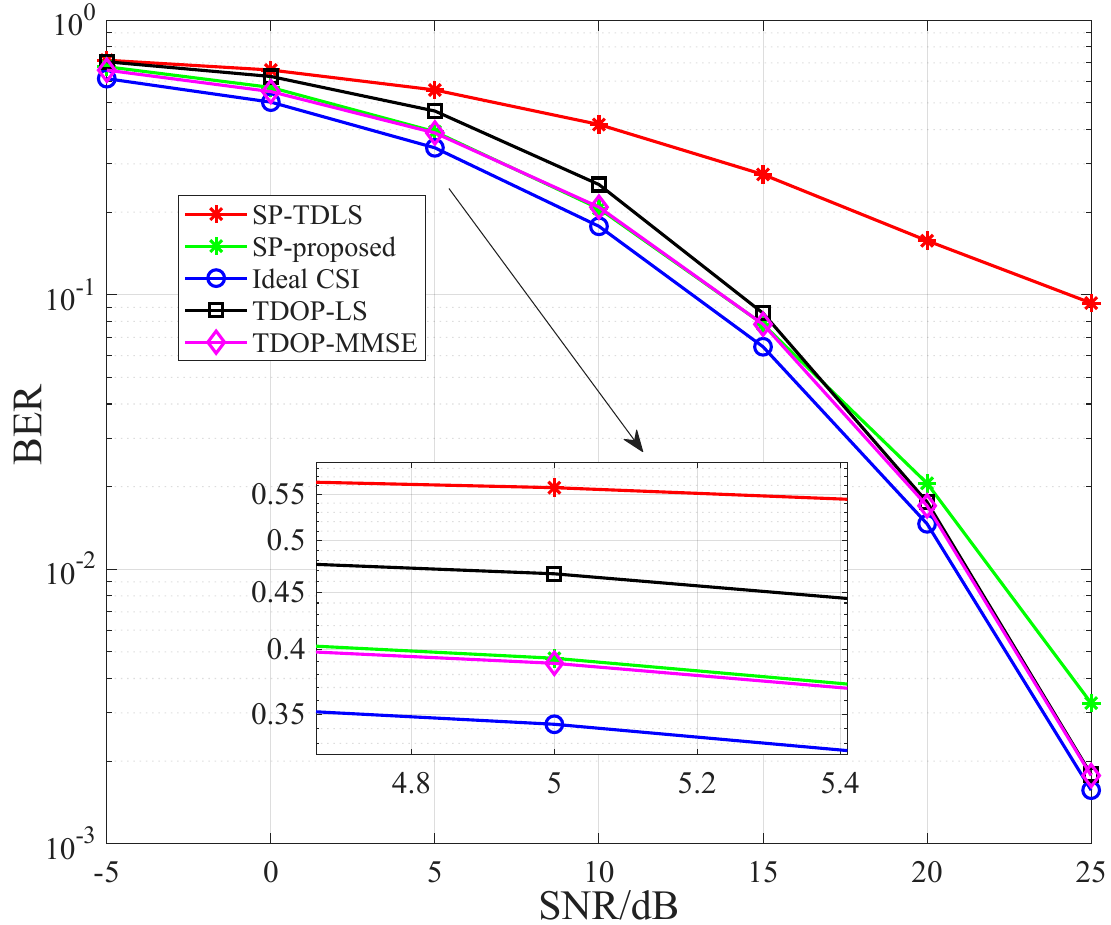}
	\caption{ BER performance with different SNR's.}
	\label{fig14}
\end{figure}

Then, the BER performance is reported with different SNR. In addition, the BER performance of traditional time-domain orthogonal pilot schemes least squares(TDOP-LS) and minimum mean square error(TDOP-MMSE) are shown as benchmarks.  For the orthogonal scheme, pilots and data signals are separated in the time domain, with a pilot overhead of $15\% $ in both time duration and power allocation. As illustrated in Fig. 14, in the low-to-medium SNR regime [-5dB, 20dB], the BER performance of the proposed scheme is comparable to that of the MMSE estimator and significantly outperforms the LS scheme. However, in the high SNR regime, the interference impact from data signal on channel estimation becomes more pronounced for the superimposed pilot approach, resulting in inferior performance compared to the orthogonal pilot scheme. However, the degradation is not prominent. Furthermore, our proposed method significantly outperforms the TDLS scheme. This advantage is attributed to the fact that we utilize both the first- and second-order statistics (mean and variance) of the received signal during channel estimation, whereas TDLS relies solely on first-order statistics.
\begin{figure}[htpb]
\includegraphics[width=7.5cm]{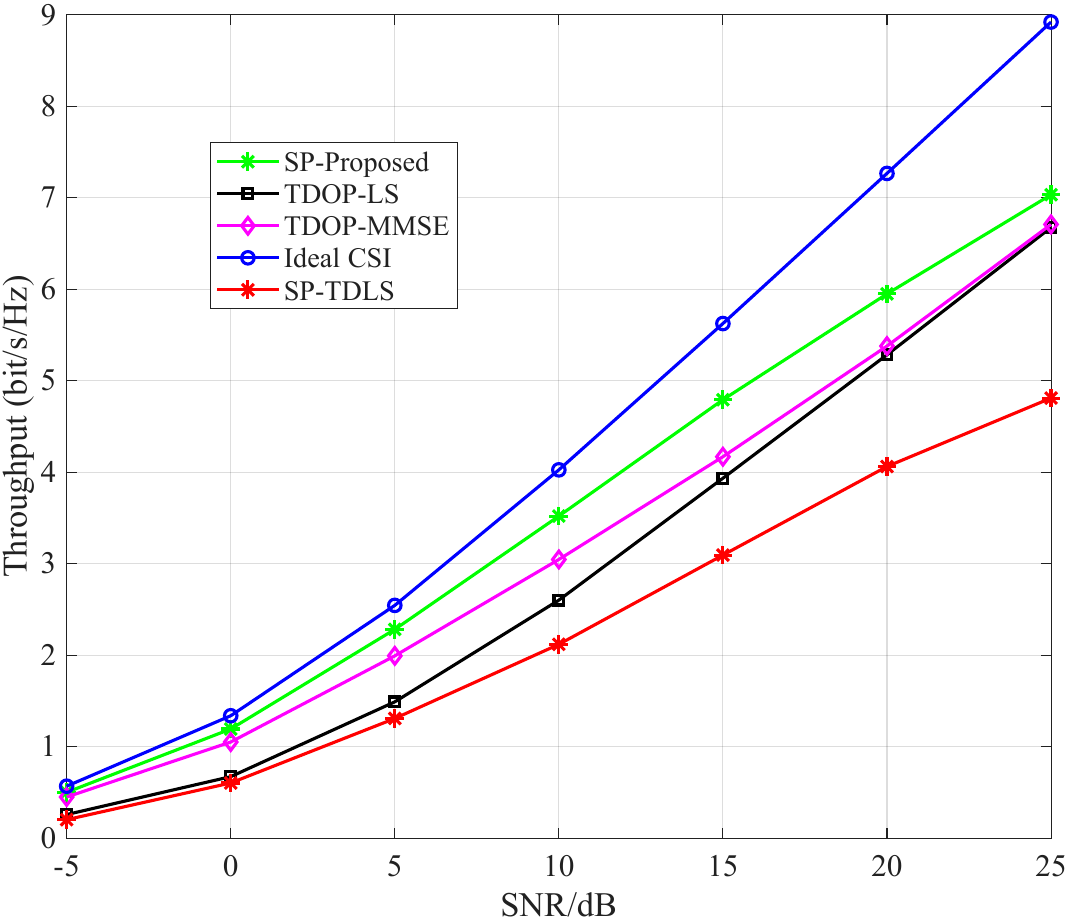}
	\caption{Throughput with different SNR's.}
	\label{fig15}
\end{figure}

Finally, we analyzed the throughput of both the superimposed and orthogonal pilot schemes, the throughput is defined as
\begin{equation}
	\mathcal{T} = \eta  \cdot {\log _2}\left( {1 + \frac{{{P_c}}}{{\sigma _n^2 + \sigma _e^2}}} \right) ,
\end{equation}
where $\eta$  denotes the resource utilization factor. Specifically, ${\eta _{sp}} = 100\% $  for the superimposed pilot scheme and ${\eta _{op}} = 85\%$  for the orthogonal pilot scheme. Additionally, ${P_c}$  denotes the signal power,  $\sigma _n^2$ is the noise power, and $\sigma _e^2$   represents the variance of the channel estimation error.

It can be observed from Fig. 15 that the proposed scheme outperforms both the traditional orthogonal pilot scheme and the superimposed pilot TDLS scheme in terms of throughput. This is attributed to the fact that the proposed scheme achieves lower channel estimation errors while maintaining a time-frequency resource utilization rate of nearly $100\%$, thereby guaranteeing superior throughput performance. In contrast, the superimposed pilot TDLS scheme exhibits significantly lower throughput, primarily due to its substantial channel estimation errors.

\begin{table*}[t]
	\centering
	\caption{PSLR and ISLR in different power weighting coefficient}
	\setlength{\arrayrulewidth}{0.4mm}
	\setlength{\tabcolsep}{16pt}
	\renewcommand{\arraystretch}{1.5}
	\begin{tabular}{c c c c c } 
		\hline
		& PSLR(Speed) & PSLR(Distance) & ISLR(Speed) & ISLR(Distance) \\ [0.5ex] 
		\hline
		$w = 0$ & -10.791 dB & -11.549 dB & -15.512 dB & -15.772 dB  \\ 

		$w = 0.2$ & -9.661 dB & -10.457 dB & -7.826 dB & -7.882 dB   \\

		$w = 0.6$ & -9.622 dB & -10.124 dB & -4.876 dB & -5.435 dB  \\
		\hline
	\end{tabular}
	
	\label{tab2}
\end{table*}
\subsection{Analysis of sensing performance}
In this section, the ambiguity function and sensing performance are analyzed. First, the distance-ambiguity and speed-ambiguity function are presented for $w = 0$, $w = 0.2$ and $w = 0.6$. The simulations are depicted in Fig. 15. Peak side lobes ratio (PSLR) and integrated sidelobes ratio (ISLR) are usually used to quantify the performance of the ambiguity function. PSLR is the ratio of the highest side lobe peak ${P_s}$ to the main lobe peak ${P_m}$, ${\rm{PSLR}} = {P_{\rm{s}}}/{P_m}$. ISLR is the ratio of the side lobe energy ${E_s}$ to the main lobe energy ${E_m}$, ${\rm{ISLR}} = {E_{\rm{s}}}/{E_m}$. The smaller PSLR and ISLR value corresponding to better side lobe leakage performance. 
\begin{figure}[htpb]
\includegraphics[width=8.5cm]{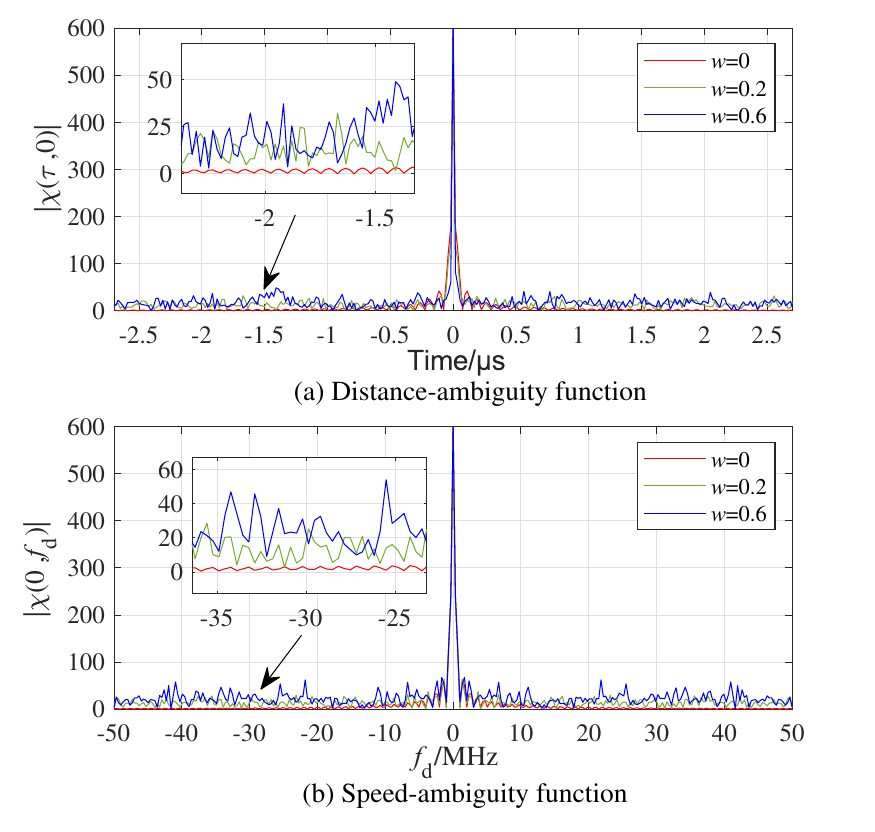}
	\caption{Ambiguity function.}
	\label{fig15}
\end{figure}
As can be seen from Fig.~\ref{fig15} and Table.~\ref{tab2}, as $w$ increases PSLR and ISLR  of the ISAC signal increases, and side lobe leakage increases, causing a decrease in sensing performance.

\begin{figure}[htpb]
\includegraphics[width=8.5cm]{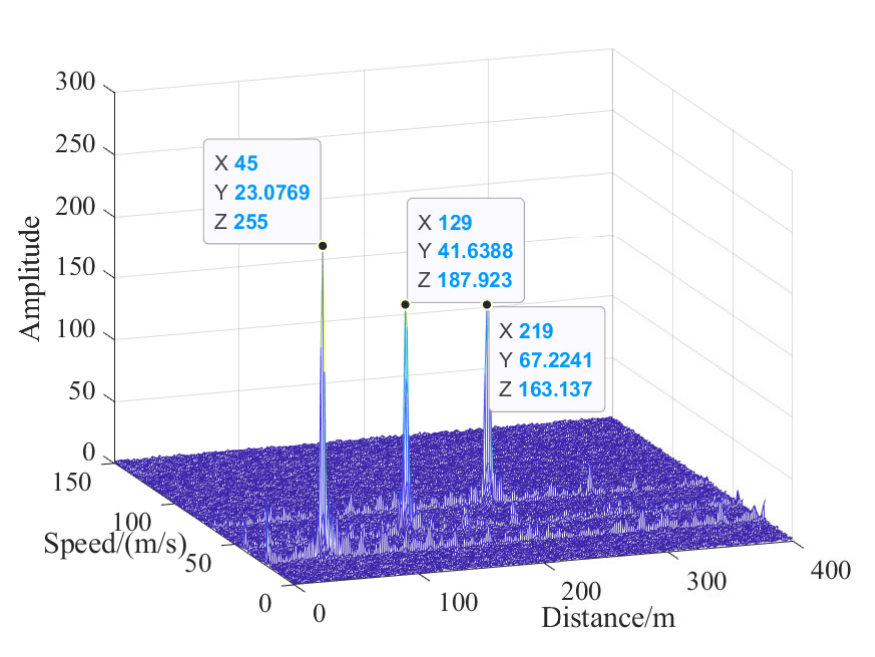}
	\caption{Simulation result of multi-target sensing.}
	\label{fig16}
\end{figure}

Next, we report the multi-target sensing performance of the proposed ISAC signal for a scenario in which the carrier frequency of 10 GHz,  bandwidth of 50 MHz, pulse width of 2 ms. We set 3 targets with the speed of ,  23.2 m/s, 41.3 m/s, 67.8 m/s and distance of 45.3 m, 130.1 m, 220.6 m, respectively.
Simulations are presented in Fig.~\ref{fig16}. The estimation errors of target speed are ${0.53\rm{\% }}$, ${0.82\rm{\% }}$, ${0.85\rm{\% }}$,while the estimation errors of target distance are ${\rm{0.67\% }}$, ${\rm{0.85\% }}$, ${\rm{0.73\% }}$. It can be known that the proposed ISAC waveform in this paper can achieve a satisfied multi-target sensing performance.

\section{Conclusion}
This paper introduced a new PTS-ISAC waveform design scheme that achieved high time-frequency resource utilization and coordination gain. Furthermore, an accurate parameter estimation method for the sensing signal was presented, which operated by separating multi-path signals. A novel approach for sensing-assisted channel estimation was presented, utilizing the cyclic maximum likelihood method and treating the sensing signal as a superimposed pilot. We analyzed the sensing performance using the ambiguity function demonstrating the capability of the proposed waveform design to achieve precise sensing. Simulations highlighted the superior accuracy and robustness of the proposed parameter estimation method. Finally, the superiority of the proposed sensing-assisted channel estimation method against conventional methods based on superimposed pilot was quantified.

\bibliographystyle{IEEEtran}
\bibliography{ref}

\end{document}